\documentclass[aps,pra,twocolumn,superscriptaddress,amsmath,amssymb]{revtex4}
\usepackage{graphicx}
\usepackage{epstopdf}
\usepackage{amsmath}
\usepackage{color}
\usepackage[normalem]{ulem}
\usepackage{graphicx}
\usepackage{subfigure}
\usepackage{float}
\usepackage{mathrsfs}
\usepackage{bbold}
\usepackage{psfrag}
\usepackage{mathcomp}
\usepackage{verbatim}
\usepackage{multirow}
\usepackage{diagbox}
\usepackage[colorlinks,citecolor=blue]{hyperref}
\def\cp#1{\mathbf{#1}}

\begin{document}

%\title{Interacting non-Hermitian ultracold atoms in three-dimensional harmonic trap: Two-body exact solutions and high-order exceptional points}
\title{Interacting non-Hermitian ultracold atoms in a harmonic trap: \\
Two-body exact solution and high-order exceptional point}
\author{Lei Pan}
\affiliation{Beijing National Laboratory for Condensed Matter Physics, Institute of Physics, Chinese Academy of Sciences, Beijing 100190, China}
\affiliation{School of Physical Sciences, University of Chinese Academy of Sciences, Beijing 100049, China}
\author{Shu Chen}
\affiliation{Beijing National Laboratory for Condensed Matter Physics, Institute of Physics, Chinese Academy of Sciences, Beijing 100190, China}
\affiliation{School of Physical Sciences, University of Chinese Academy of Sciences, Beijing 100049, China}
\affiliation{The Yangtze River Delta Physics Research Center, Liyang, Jiangsu 213300, China}
\author{Xiaoling Cui}
\email{xlcui@iphy.ac.cn}
\affiliation{Beijing National Laboratory for Condensed Matter Physics, Institute of Physics, Chinese Academy of Sciences, Beijing 100190, China}
\affiliation{Songshan Lake Materials Laboratory , Dongguan, Guangdong 523808, China}
\date{\today}

\begin{abstract}
We study interacting ultracold atoms in a three-dimensional (3D) harmonic trap with spin-selective dissipations, which can be effectively described by  non-Hermitian parity-time ($\mathcal{PT}$) symmetric Hamiltonians.
%In our previous work [Phys. Rev. A {\bf 99}, 011601(R)(2019)], we have shown that a two-species(spin-1/2) one-dimensional(1D) Bose gas in such setup can exhibit high-order exceptional points(EPs) and create ultra-sensitive spectral response via interaction anisotropies in spin channels.
%In this paper, we extend the results to three-dimensional(3D) atomic systems with high spins.
%We generalize the exact solution of two atoms in harmonic trap with s-wave interaction to arbitrary spin systems with the non-Hermitian potential.
By solving the non-Hermitian two-body problem of spin-1/2 (spin-1) bosons in a 3D harmonic trap exactly, we find that the system can exhibit third-order  (fifth-order) exceptional point (EP) with ultra-sensitive cube-root (fifth-root) spectral response due to interaction anisotropies in spin channels.
%can create an ultra-sensitive cube-root (fifth-root) spectral response $\sim \epsilon^{1/3}$($\sim \epsilon^{1/5}$) to tiny perturbation $\epsilon$ introduced by interaction anisotropies in spin channels.
We also present the general principle for the creation of high-order EPs and their spectral sensitivities with arbitrary particle number $N$ and arbitrary spin $s$. Generally, with spin-independent interactions, the EP order of bosons can be as high as $2Ns+1$, and the spectral response around EP can be as sensitive as $\sim \epsilon^{1/(2ks+1)}$ under a $k$-body interaction anisotropy. Moreover, we propose to detect the ultra-sensitive spectral response through the probability dynamics of certain state.  These results suggest a convenient route towards more powerful sensor devices in spinor cold atomic systems.
\end{abstract}
\maketitle

\section{Introduction}
%Conventional quantum theory demands that a Hamiltonian(or other operator representing an observable quantity) must be Hermitian which ensures that the eigenvalues of a Hamiltonian are real and bounded below. (widely existing)
%Dissipation is a ubiquitous phenomena in nature since the physical systems are always inevitable to interact with environment.
An open system with dissipative processes can be described by a non-Hermitian Hamiltonian phenomenologically. Among various types of non-Hermitian Hamiltonians, the parity-time($\mathcal{PT}$) symmetric Hamiltonian\cite{Bender} is a peculiarly fascinating one whose spectra can be real and bounded below.
%Hamiltonians that are non-Hermitian have traditionally been used to describe the open systems with dissipative processes. In their pioneering work, Bender and Boettcher found that there exists a kind of non-Hermitian systems whose spectra are also real and bounded below whose spectra are also real and bounded below are known as PT-symmetric systems.
With tunable parameters, such systems can undergo a spontaneous $\mathcal{PT}$-symmetry breaking transition, where the eigenvalues of the system start to develop imaginary parts. % complex
Right at the transition, two or more eigenvalues and their corresponding eigenvectors coalesce simultaneously, and the location is known as the exceptional point(EP)\cite{Heiss1,Moiseyev,Kato}.
Different from conventional degenerate point(DP) in Hermitian systems, EP in non-Hermitian systems can exhibit an ultra-sensitive response to external perturbations. Specifically,
%For the conventional DPs in Hermitian systems, a asymmetric perturbation would lead to an energy splitting which is the same order as the perturbation itself $\sim \epsilon$ at most. While
around an EP of $q$-th order, where $q$ is the number of eigenvectors that simultaneously coalesce, a small perturbation of strength $\epsilon$ can result in a large energy splitting $\sim \epsilon^{1/q}$. In comparison, near a conventional DP, any perturbation can at most give rise to a linear energy shift $\sim \epsilon$.

Given above properties, there have been a growing recognition that non-Hermitian EP systems can be an ideal candidate for making sensors\cite{Wiersig1,Fleury,Wiersig2,Liu,Ding,Yang}, and those with high-order EPs are particularly  attractive given their growing sensitivity.  % with the order of EP as high as possible.    EPs and in particular high-order EPs, have been gradually in non-Hermitian systems can provide a new pathway of enhancing their sensitivity beyond the standard DPs arrangements in Hermitian systems and such sensitive response to tiny perturbations makes the non-Hermitian EPs system an ideal candidate for sensors\cite{Wiersig1,Fleury,Wiersig2,Liu,Ding,Yang}.
In the past few years, experiments on various photonic, acoustic and atomic systems have realized the second-order ($q=2$) EPs\cite{Dembowski,Dietz,Lee,Choi,Guo,Lin1,Feng1,Zhen,Sun,Doppler,Xu,Dembowski2,Yang2,Miao,Gao,Hodaei1,Feng2,Ruter,Regensburger,Liertzer,Zhu,Brandstetter}, and later third-order ($q=3$) \cite{Ding2,Hodaei2} and even higher-order ones ($q>3$)\cite{Wang}. Moreover, the ultra-sensitive spectral responses have been successfully detected near third-order EPs\cite{Hodaei2}. %{\color{red} (why this reference is not cited previously together with ref.33? ; ref.34 also detect spectral sensitivity? pls check) }.
%In principle, even higher-order EPs(greater than second order)should further amplify the effect of perturbations, leading to even greater sensitivity but such higher-order EPs are more difficult to realize in laboratories while such
Theoretically, higher-order EPs have also been proposed by a number of studies in literature\cite{Graefe,Demange,Teimourpour,Heiss2,Heiss3,Lin,Jing,Schnabel,Zhong}.
%Very recently, two groundbreaking experiments have successfully achieved the third-order EPs and detected the enhanced sensitivity in coupled acoustic cavities\cite{Ding2} and optical micro-ring system\cite{Hodaei2}. However, such higher-order EPs and the accompanying ultrasensitive responses in ultracold atomic arrangements have yet to be observed.

\begin{figure}[ht]
\includegraphics[width=8.5cm]{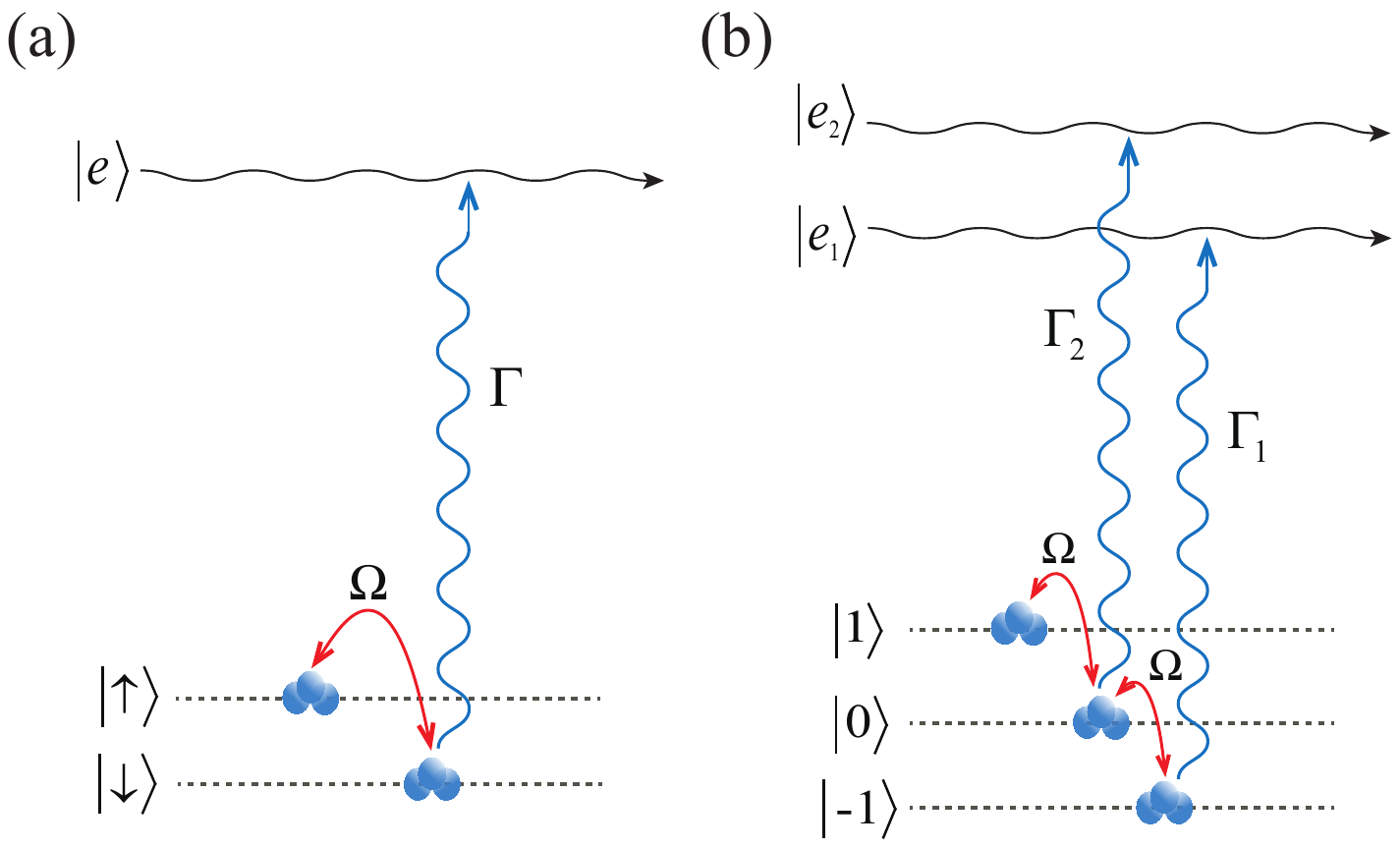}
\caption{(Color online).
Schematics of experimental set-up for ultracold atoms with spin-selective dissipations. (a) Two-species (spin-1/2) system where an rf field is used to couple the two-spin states with coupling strength $\Omega$. A  resonant laser is applied to generate a spin-dependent dissipation ($\Gamma$) on spin-down state. (b) Three-species (spin-1/2) system where two rf fields couple the three spin states and two additional lasers with different strengths ($\Gamma_1$, $\Gamma_2$) are used to transfer the states $|0\rangle$, $|-1\rangle$ to excited atomic states $|e_1\rangle$, $|e_2\rangle$.} \label{fig1}
\end{figure}

%In recent years, ultracold atomic gases have emerged as an ideal platform for quantum simulation in view of their high controllability. Very recently,

In ultracold atoms, by using the laser-induced spin-selective dissipation, a non-Hermitian atomic gas possessing $\mathcal{PT}$-symmetry has become accessible in experiments\cite{Luo}. Fig.\ref{fig1} shows the schematics for such setup in two-species (spin-1/2) and three-species (spin-1) atomic systems. Specifically, for spin-1/2 system ($\uparrow,\downarrow$), a laser field uniquely transfers the spin-down atom to a highly excited atomic state and causes losses only in this spin state. Such spin-dependent losses can be described by an imaginary magnetic field, $i\Gamma s_z$, up to a constant energy shift ($\sim -i\Gamma/2$). Together with an additional radio-frequency(rf) field (with strength $\Omega$) to couple $\uparrow$ and $\downarrow$ states, this setup realizes the following type of %$\mathcal{PT}$-symmetric
Hamiltonian:% in spin space:
\begin{equation}
h_{PT}=\Omega s_x+i\Gamma s_z, \label{H_PT}
\end{equation}
which supports a second-order EP at $\Omega=\Gamma$. The Hamiltonian (\ref{H_PT}) is invariant under the $\mathcal{PT}$ symmetry transformation, where the parity operator $\mathcal{P}$ can be represented by the standard involutory permutation matrix and the time reversal operator $\mathcal{T}$ is equivalent to complex conjugation.
%Here the parity $\mathcal{P}$ is represented by the standard involutory permutation matrix and the time-reversal operator $\mathcal{T}$ is the complex conjugation.
Similarly, for spin-1 atoms, apart from the rf fields, two additional lasers with different strengths ($\Gamma_1, \Gamma_2$) can be applied to transfer two of the spin states to excited atomic states. When tuning the relative dissipation strengths of two lasers to be $\Gamma_1/2=\Gamma_2=\Gamma$, one can realize the same $\mathcal{PT}$-symmetric Hamiltonian as Eq.\ref{H_PT} with $s_{\alpha}$ substituted by the spin-1 operators.
%one can equally realize the  $\mathcal{PT}$-symmetric Hamiltonian as Eq.\ref{H_PT}, where $s_{\alpha}$ becomes the spin-1 operators.

Given the above $\mathcal{PT}$-symmetric potential potential for a single atom, it is natural and interesting to investigate the  interplay of such non-Hermitian potential and the highly-tunable interactions in cold atoms? In particular, is it possible to utilize an interacting ultracold atomic gas for designing sensors? To answer these questions, in our previous work\cite{Pan}, we have investigated the repulsively interacting 1D spin-1/2 Bose gas with $\mathcal{PT}$ potentials, and found that such system can indeed be used to generate arbitrarily high-order EPs and produce ultra-sensitive spectral response through interaction anisotropies in spin channels. This is facilitated by the intrinsic ferromagnetic correlation in such system\cite{Li,Guan}.  In the present work, we consider the more general 3D atomic systems with high spin, in which it is easier to achieve higher order EPs than that in spin-1/2 systems. We exactly solve the two-body problems of spin-1/2 and spin-1 bosons under $\mathcal{PT}$ potential in a 3D harmonic trap, from which we establish the mean-field treatment for weak coupling bosons in the repulsive scattering branch. Using the mean-field treatment, we further study the order of EPs and their associated spectral sensitivity against interaction anisotropies for a small cluster of spin-1/2 and spin-1 bosons, and finally extend to many bosons with arbitrary spin.
%We present general rules for creating high-order EPs in these systems and give the
In general, we show that for a $N$-particle system with spin-$s$ bosons, %for spin-$s$ bosons with particle number $N$
the EP order can be as high as $2Ns+1$ with spin-independent interactions, and under a tiny $k$-body interaction anisotropy (with strength $\epsilon$), the spectral splitting around the EP sensitively scales as $\sim \epsilon^{1/(2ks+1)}$. %To be more concrete, take spin-1 bosons for instance, two(three) bosons under a two(three)-body interaction anisotropy can be used to create a fifth(seventh)-order EP and a spectral shift as sensitive as $\sim \epsilon^{1/5}$ ($\sim \epsilon^{1/7}$). %In addition, in this work we also discuss the generalization to high-spin fermions.
These results could serve as a guideline for designing powerful sensor devices in spinor cold atoms systems.

The rest of the paper is organized as follows. In Sec.\ref{sec2} we present the formalism of solving the non-Hermitian two-body problem in a harmonic trap with an arbitrary spin.
In Sec.\ref{sec3}, we apply the two-body exact solution to spin-1/2 and spin-1 bosons, and discuss the spectral response with respect to two-body interaction anisotropies. %Here we will also demonstrate the validity of mean-field treatment to repulsive scattering branch in weak coupling limit.
Sec.\ref{sec4} is contributed to the spectral sensitivity of three  spin-1/2 and spin-1 bosons against interaction anisotropies in both two-body and three-body collision sectors. % using the established mean-field treatment in weak coupling regime of scattering branch.
In Sec.\ref{sec5}, we present the mathematical origin for the order of EPs and their associated sensitivities, and generalize the rules to many-body systems with an arbitrary spin. Sec.\ref{sec6} contributes to experimental detection of ultra-sensitive spectral response through dynamics. Finally we conclude in Sec.\ref{sec7}.

\section{Formalism for two-body problem in trapped non-Hermitian system with an arbitrary spin} \label{sec2}

In this section, we study the two-body problem of $s$-wave interacting cold atoms in a 3D harmonic trap, with non-Hermitian external potential and with arbitrary spin. The two-body problem in trapped Hermitian system has been exactly solved in Ref.\cite{Busch}. Here, the two-body  system can be described by $H=H_0+U$, where (we set $\hbar=1$ throughout the paper)
\begin{eqnarray}
H_0&=&\left(-\frac{1}{2m}\mathbf{\nabla}_1^2-\frac{1}{2m}\mathbf{\nabla}_2^2+\frac{1}{2}m\omega_{T}^2{\cp r}_1^2+\frac{1}{2}m\omega_{T}^2{\cp r}_2^2\right) \nonumber \\
&&+\Omega(s_{1x}+s_{2x})+i\Gamma(s_{1z}+s_{2z}); \nonumber\\
U&=&\sum_{S,M}g_S^MP_S^M\delta(\mathbf{r})
\label{Ham_2b}
\end{eqnarray}
here ${\cp r}={\cp r}_1-{\cp r}_2$ is the relative coordinate of two atoms; $s_{i\alpha}\ (\alpha=x,y,z)$ denotes the spin-$s$ operator for the $i$-th atom; $g_S^M$ is the bare coupling in the scattering channel with total spin $S$ and total magnetization $M$, and $P_S^M$ is the corresponding projection operator; $g_S^M$ can be related to the s-wave scattering length  $a_{s}^{SM}$ via the renormalization equation:
\begin{equation}
\frac{1}{g_S^M}=\frac{m}{4\pi a_{s}^{SM}} - \frac{1}{V}\sum_{\cp k} \frac{m}{k^2}. \label{renor}
\end{equation}
Since the center-of-mass motion (related to coordinate ${\cp R}=({\cp r}_1+{\cp r}_2$)/2) can be decoupled from the problem, from now on we only focus on the relative motion (related to ${\cp r}$) of two atoms and solve the Schr{\" o}dinger equation
\begin{equation}
(H_{0}^{rel}+U)|\Psi\rangle=E_{rel}|\Psi\rangle.
\end{equation}
For the bound state, $|\Psi\rangle$ satisfies the Lippman-Schwinger equation
\begin{equation}
\begin{split}
|\Psi\rangle=G_EU|\Psi\rangle  \label{psi}
\end{split}
\end{equation}
where $G_E=(E_{rel}-H_{0}^{rel})^{-1}$ is Green function. Due to the conservation of total spin $S$, one can solve the two-body problem in each $S$-sector individually. Specifically, we introduce a set of variables $\{f_{M}\}$ to express					 %the scattering amplitudes in different channels:
\begin{equation}
\langle {\cp r}|U|\Psi\rangle=\sum_{M=-S}^S f_M|SM\rangle \delta({\cp r}), \label{U1}
\end{equation}
with $|SM\rangle$ the two-spin state with total spin $S$ and total magnetization $M$. By plugging (\ref{U1}) into (\ref{psi}), we arrive at $2S+1$ coupled equations for $\{ f_M\}$, which lead to a non-trivial solution only under the condition
\begin{equation}
{\rm Det}\left( \frac{1}{g_S^M}\delta_{MM'} -\langle M| G_E(0,0) |M'\rangle \right) =0.  \label{E}
\end{equation}
This is a $(2S+1)\times(2S+1)$ matrix equation, from which one can obtain the bound state energy $E_{rel}$. Here the Green function can be expanded as
\begin{equation}
G_E({\cp r},{\cp r'})=\sum_n \sum_j \frac{\psi_n({\cp r})\psi^*_n({\cp r}')}{E_{rel}-E_n-\epsilon_j} \frac{|\mu_j^R\rangle \langle \mu_j^L|}{\langle \mu_j^L|\mu_j^R\rangle},  \label{G}
\end{equation}
where $\psi_n({\cp r})$ and $E_n$ are respectively the eigen-wavefunction and eigen-energy of the $n$-th harmonic oscillator level; $|\mu_j^R\rangle$ and $|\mu_j^L\rangle$ are the right and left spin vectors in the total spin $S$ subspace, which are defined through $H_{PT}|\mu_j^R\rangle=\epsilon_j|\mu_j^R\rangle$ and $H_{PT}^\dagger|\mu_j^L\rangle=\epsilon_j^*|\mu_j^L\rangle$\cite{footnote}, here $H_{PT}=\Omega(s_{1x}+s_{2x})+i\Gamma(s_{1z}+s_{2z})$. The corresponding bra is defined by $\langle\mu_i^{L,R}|=(|\mu_i^{L,R}\rangle)^\dagger$. Since the spin expansion in (\ref{G}) fails at the location of single-particle EP ($\Gamma=\Omega$), one needs to resort to the exact diagonalization to obtain the spectrum at EP.

\section{Two-body exact solutions of spin-1/2 and spin-1 bosonic systems} \label{sec3}

In this section, we present the exact solutions for two-boson system in a 3D harmonic trap, with both spin-1/2 (two species) and spin-1(three species), following the formalism in Sec.\ref{sec2}. We will then focus on the spectral response in the weak coupling regime of repulsive scattering branch, where the mean-field treatment can be justified by exact solutions.

\subsection{Spin-1/2}

%The spin-1/2 $\mathcal{PT}$ system can be realized in experiment and the schematic diagram of experimental set-up is shown in Fig.\ref{fig1}(a).
Previously, we have solved the two-body exact solutions of 1D bosons with $\mathcal{PT}$ potential\cite{Pan}. In the 3D case, since bosons are still scattering in total spin $S=1$ channel regardless of dimension, one again needs to solve a $3\times 3$ matrix equation as shown by Eq.\ref{E}. However, different from 1D case, in 3D the bare coupling part has an ultraviolet divergence at high energy (see Eq.\ref{renor}), which should be cancelled exactly by the same divergence in the Green function part (see Eq.\ref{G}).  To facilitate the presentation of results, we define the confinement length $l=1/\sqrt{m\omega_T}$ as a typical length scale. %, and use the trapping frequency $\omega$ as a typical energy scale.

\begin{figure}[h]
\includegraphics[width=8.3cm]{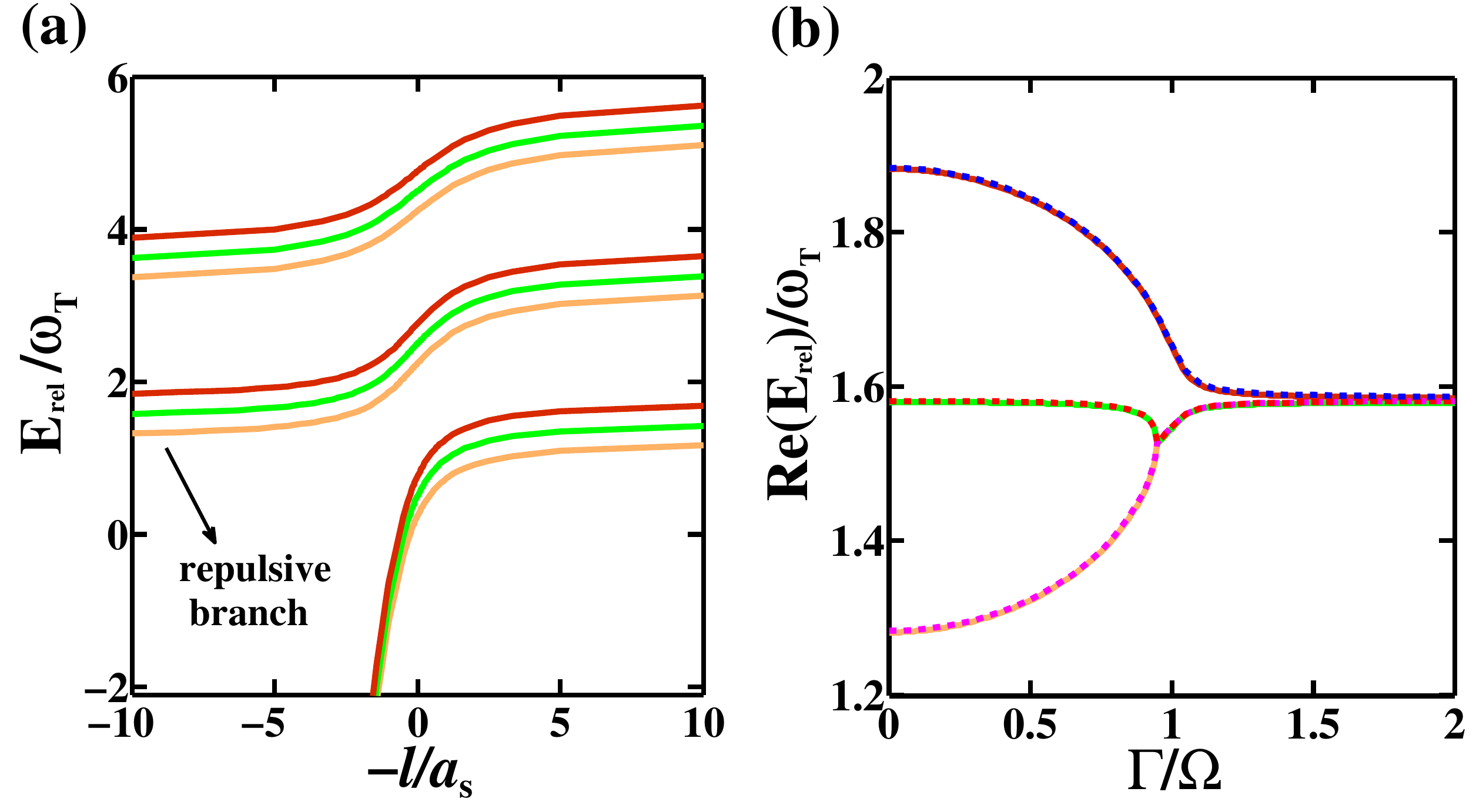}
\includegraphics[width=8.5cm]{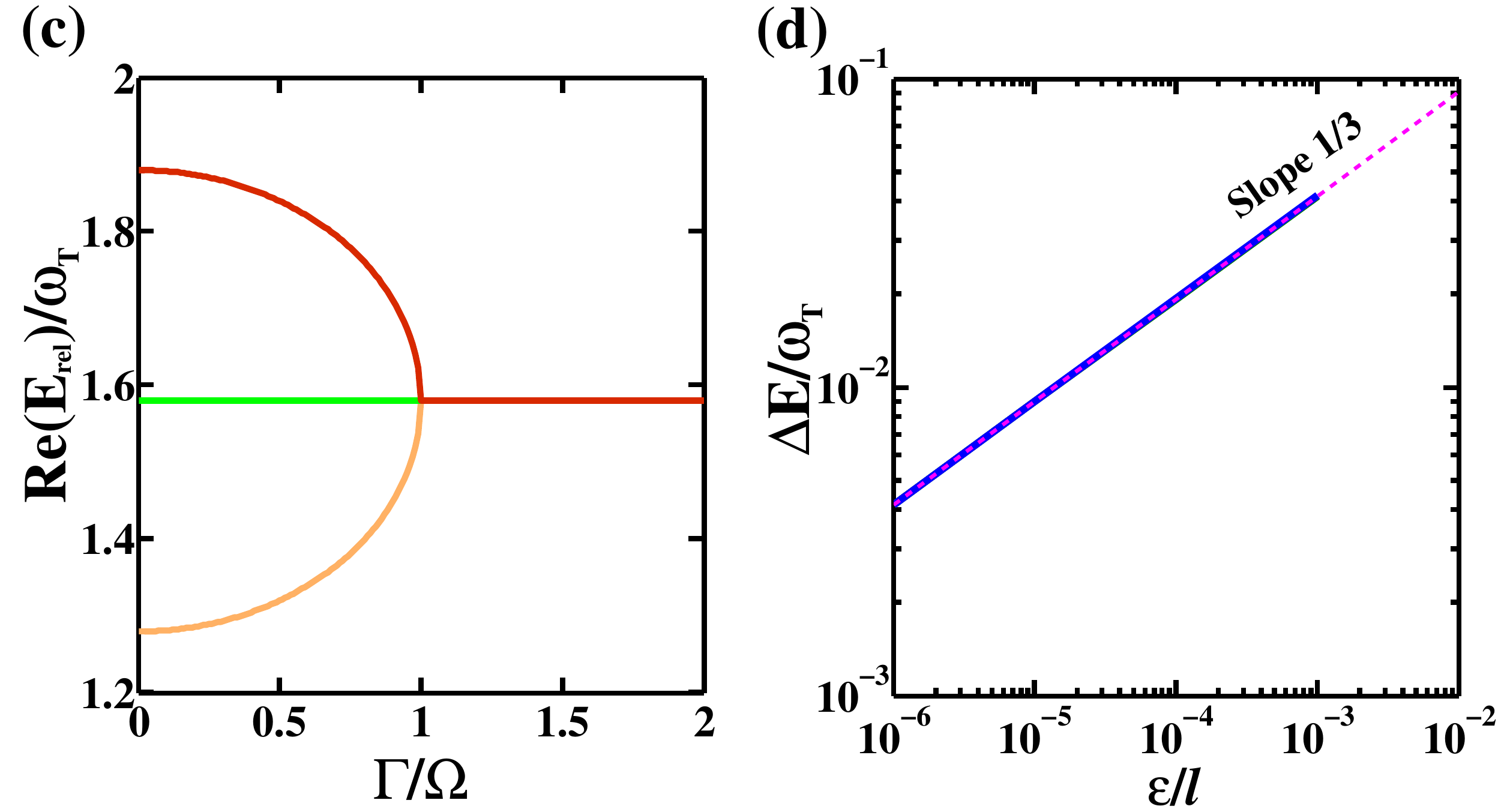}
\caption{(Color online). Exact spectrum of non-Hermitian two spin-1/2 boson system in a 3D harmonic trap. (a) Energy levels as functions of coupling strength. Here we take $a_s^{1,1}=a_s^{1,-1}\equiv a_s$ and $a_s^{1,0}=a_s+\delta$ with $\delta=0.005l$; $\Omega/\omega_T=0.3$ and $\Gamma/\Omega=0.5$. The location of repulsive branch in weak coupling regime ($0<a_s/l \ll 1$) is marked. (b)Exact solutions of the three lowest energy levels at repulsive branch as functions of $\Gamma$, in comparison with the mean-field prediction (dashed lines). Here we take $a_s^{1,\pm1}=a_s=0.1l$ and $a_s^{1,0}=a_s+\delta$ with $\delta=0.005l$. (c) The same as (b) except for $\delta=0$, i.e., when the interaction is spin-independent. In this case, $\Gamma=\Omega$ is the location of third-order EP. (d) Energy splitting($\Delta E$) at EP as a function of interaction anisotropy $\epsilon$ in $\{S=1,M=0\}$ channel. The dashed line in shows analytical fit (see text). \\
%{\color{red} $E$ change to $E_{rel}$ in all figures, if it is only the energy for relative motion}
} \label{fig2}
\end{figure}

The two-body results are shown in Fig.\ref{fig2}. In Fig.\ref{fig2}(a), we plot the exact spectrum as a function of interaction strength, given fixed $\Omega$, $\Gamma$ and a small interaction anisotropy in $S=1,M=0$ scattering channel: $a_s^{1,1}=a_s^{1,-1}\equiv a_s$ and $a_s^{1,0}=a_s+\delta$ with $\delta=0.005l$. The weak coupling regime of repulsive scattering branch, which is located at $0<a_s/l \ll 1$, is also marked in the plot. As seen from Fig.\ref{fig2}(b), the exact solutions of the lowest three levels in this regime (solid lines) can be well predicted by the mean-field energy shift based on the non-interacting ground state  (dashed line).

In Fig.\ref{fig2}(c), we plot the spectrum of the lowest three levels as functions of $\Gamma$ given a spin-independent interaction ($a_s^{1,1}=a_s^{1,-1}=a_s^{1,0}\equiv a_s$). In this case, since the $\mathcal{PT}$ potential commutes with the interacting Hamiltonian, the EP still occurs at $\Gamma=\Omega$, same as the non-interacting system. However, different from the single-particle picture, here the EP order is upgraded to three, instead of two, simply because the two-boson scattering is locked in the spin-triplet channel. As shown in Fig.\ref{fig2}(c), three energy levels coalesce simultaneously  at this point, and we have checked that the three eigenstates also coalesce at this point, thus featuring a third-order EP. In comparison to the presence of third-order EP in 1D case\cite{Pan}, we remark that they share the same physical origin, i.e.,  the bosonic statistics requires two bosons scattering in the triplet (ferromagnetic) channel.

Comparing Fig.\ref{fig2}(b) with (c), one can see that a small interaction anisotropy can completely destroy the triple degeneracy at EP. The interaction effect on the shift of degeneracy point is studied recently in Ref.\cite{Yu}. Here we focus on the sensitive change of spectrum near EP, which is essential for the implement of sensor devices. For the weak coupling regime, this can be analyzed efficiently in a mean-field manner.. Given a small anisotropy in $M=0$ scattering channel, i.e.,  $a_s^{1,1}=a_s^{1,-1}\equiv a_s$ and   $a_s^{1,0}=a_s+\epsilon$, we can expand the mean-field Hamiltonian in the space of three triplet states ($\{M=1,0,-1\}$) as \begin{equation}
\begin{split}
H_{MF}=\frac{4\pi a_s}{m}|\psi_0(0)|^2\Bbb1+
\left( \begin{array}{cccc}
   i\Gamma & \frac{\sqrt{2}}{2}\Omega & 0 \\
   \frac{\sqrt{2}}{2}\Omega & \frac{4\pi}{m}|\psi_0(0)|^2\epsilon & \frac{\sqrt{2}}{2}\Omega \\
   0 & \frac{\sqrt{2}}{2}\Omega & -i\Gamma
\end{array}
\right)
\label{H_3}
\end{split}
\end{equation}
Where $\Bbb1$ is the identity matrix and $\psi_0({\cp r})$ denotes the ground-state wave-function in 3D harmonic trap. Diagonalizing (\ref{H_3}) at EP($\Gamma=\Omega$) for small $\epsilon$, we obtain the three energy shifts as $\Delta E_1=\Delta E$, $\Delta E_2=\Delta E\exp(i\frac{2\pi}{3}) $, and $\Delta E_3=\Delta E \exp(i\frac{4\pi}{3})$, which have the same amplitude
\begin{equation}
\Delta E=\left(\frac{4\pi\Omega^2}{m}|\psi_0(0)|^2 \epsilon\right)^{\frac{1}{3}}.\label{E_10}
\end{equation}
We can see that this expression matches well with the energy shift from exact numerical calculations, as shown in Fig.\ref{fig2}(d). The cube-root dependence of $\Delta E$ on the perturbation parameter $\epsilon$ is a deterministic feature of the third-order EP.

\subsection{Spin-1}\label{spin1}

Spin-1 bosons can scatter in total spin $S=2$ and $S=0$ channels, which are respectively associated with scattering lengths $a_s^{S=2}\equiv a_2$ and $a_s^{S=0}\equiv a_0$. Depending on the relative value of $a_2$ and $a_0$, the ground state of the system can show different magnetic orders\cite{spin1_H_1,spin1_H_2}. For $a_2<a_0$, the ground state is ferromagnetic and the typical atomic system is $^{87}$Rb; while for $a_0<a_2$, the ground state is anti-ferromagnetic (spin-singlet) and the typical atomic system is $^{23}$Na. In this section, we will show that depending on the magnetic order or the ground state scattering channel of bosons, the $\mathcal{PT}$ potential can exhibit rather distinct effects.

To facilitate discussions, we rewrite the $\mathcal{PT}$ potential as
\begin{equation}
H_{PT}=\Omega S_x+i\Gamma S_z, \label{PT2}
\end{equation}
with $S_{\alpha}=\sum_i s_{i,\alpha}$ the $\alpha(=x,y,z)$ component of total spin operator. Since  $H_{PT}$ commutes with total spin ${\cp S}^2$,  it will not couple states with different $S$ but just induced coupling within the same $S$ between different $M$. If the ground state is spin singlet $|S=0\rangle$, then the $\mathcal{PT}$ potential will take no effect because $H_{PT}|S=0\rangle=0$. This means that for bosons with anti-ferromagnetic order, such as $^{23}$Na, the ground state will not be affected by $\mathcal{PT}$ potential. In comparison, for bosons with ferromagnetic order, such as $^{87}$Rb, $H_{PT}$ can take dramatic effect. In the latter case, one needs to solve a $5\times 5$ matrix equation expanded in $\{S=2,M=\pm 2,\pm 1,0\}$ subspace, as shown by Eq.\ref{E}.

\begin{figure}[h]
\includegraphics[width=8.3cm]{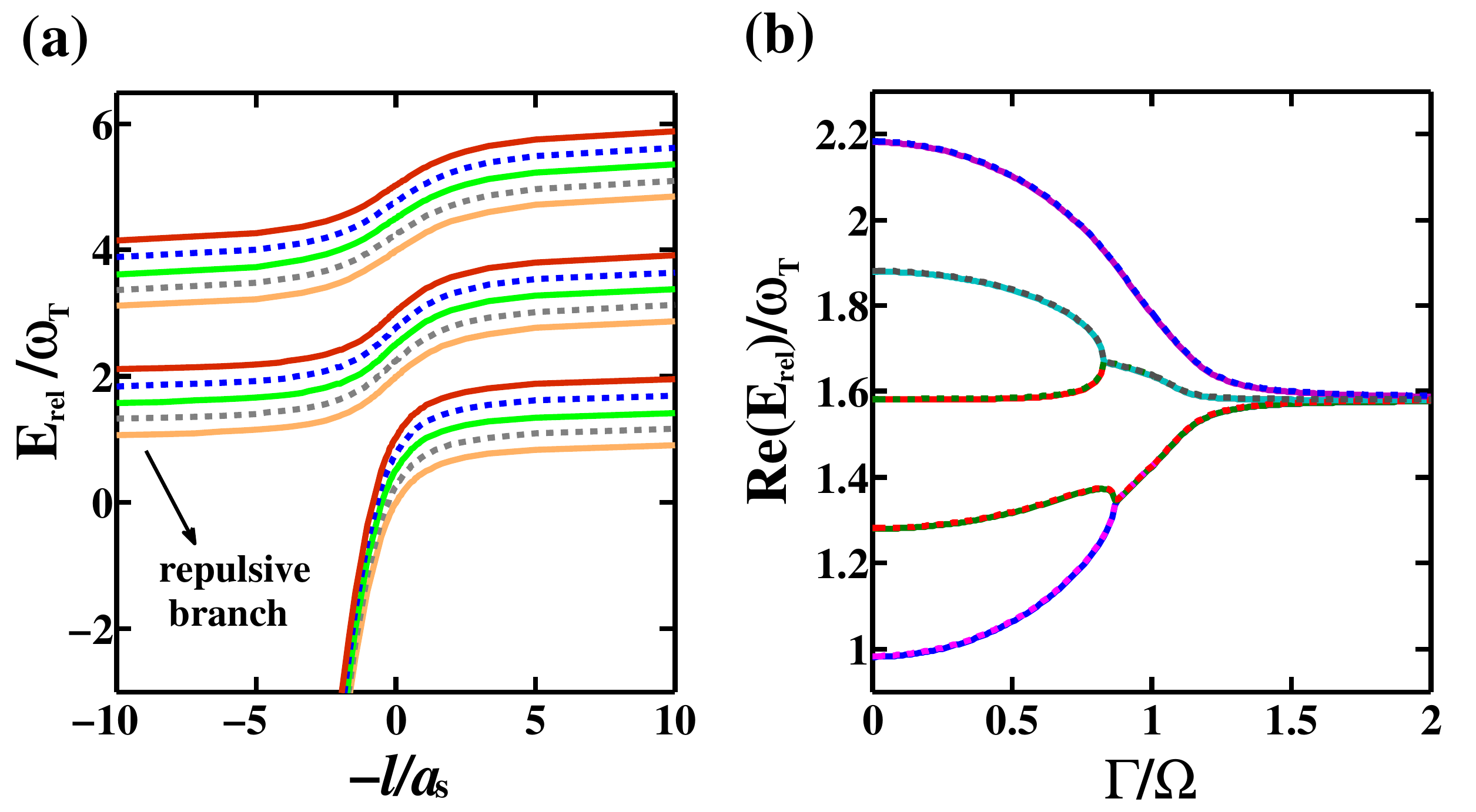}
\includegraphics[width=8.5cm]{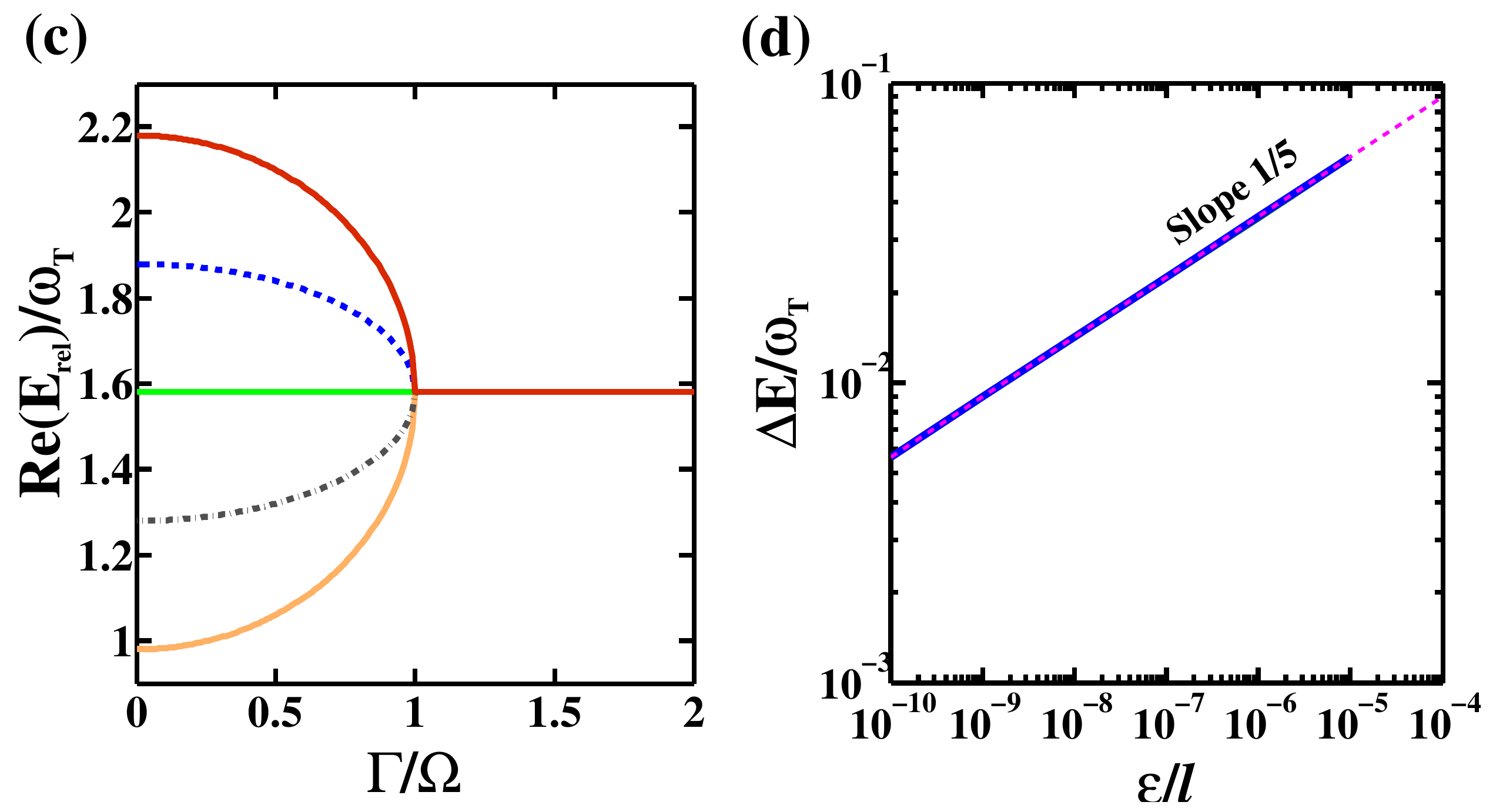}
\caption{(Color online).
Exact spectrum of non-Hermitian two spin-1 boson system in a 3D harmonic trap. (a) Energy levels as functions of coupling strength. Here we take $a_s^{2,2}=a_s^{2,1}=a_s^{2,-1}=a_s^{2,-2}\equiv a_s$ and $a_s^{2,0}=a_s+\delta$ with $\delta=0.005l$; $\Omega/\omega=0.3$ and $\Gamma/\Omega=0.5$. The location of repulsive branch in weak coupling regime ($0<a_s/l \ll 1$) is marked. (b)Exact solutions of the five lowest energy levels at repulsive branch as functions of $\Gamma$, in comparison with the mean-field prediction (dashed lines). Here we take $a_s^{2,\pm2}=a_s^{2,\pm1}=a_s=0.1l$ and $a_s^{2,0}=a_s+\delta$ with $\delta=0.005l$. (c) The same as (b) except for $\delta=0$, i.e., when the interaction is spin-independent. In this case, $\Gamma=\Omega$ is the location of fifth-order EP. (d) Energy splitting($\Delta E$) at EP as a function of interaction anisotropy $\epsilon$ in $\{S=2,M=0\}$ channel. The dashed line shows analytical fit (see text).} \label{fig3}
\end{figure}

In Fig.\ref{fig3}(a), we plot the lowest five energy levels for two bosons in $S=2$ sector with an anisotropic interaction in $M=0$ channel. In weak coupling regime, the spectrum in the repulsive scattering branch can be well predicted by mean-field theory, see Fig.\ref{fig3}(b). Namely, we expand
%Fig.\ref{fig3}(b)  (see (a)), in comparison to those with a small interaction anisotropy in $S=2,M=0$ channel (see (b)). We see that for isotropic interaction, i.e., all $a_s^{SM}=a_s$, the five energy levels (as well as the associated eigenstates) merge at EP ($\Gamma/\Omega=1$), which marks the location of a fifth-order EP. Near this super-degenerate point, the energy spectrum can be sensitively influenced by a small interaction anisotropy, as shown in (b). Here we take $a_s^{2,M}=a_s=0.1l$ for $M=\pm 2,\pm1$ and $=a_s+\epsilon$ for $M=0$ with $\epsilon=0.005l$. In Fig.\ref{fig3}, we plot the amplitude of energy shifts in terms of the interaction anisotropy $\epsilon$, which indeed shows a fifth-root dependence $\Delta E\sim \epsilon^{1/5}$. To understand this, we again use the mean-field theory and expand
the Hamiltonian in the subspace of $S=2$ sector, which reads (up to a constant mean-field shift $\frac{4\pi a_s}{m}|\psi_0(0)|^2$):
\begin{equation}
\begin{split}
H_{MF}=
\left( \begin{array}{cccccc}
  2i\Gamma & \Omega & 0 & 0 & 0\\
  \Omega  & i\Gamma  & \frac{\sqrt{6}}{2}\Omega & 0 & 0\\
  0  & \frac{\sqrt{6}}{2}\Omega & \frac{4\pi}{m}|\psi_0(0)|^2\epsilon & \frac{\sqrt{6}}{2}\Omega & 0\\
  0  & 0 & \frac{\sqrt{6}}{2}\Omega & -i\Gamma  & \Omega\\
  0  & 0 & 0 & \Omega & -2i\Gamma
\end{array}
\right)
\label{H_5}
\end{split}
\end{equation}
For small $\epsilon$, one can easily obtain the energy shifts of five levels at $\Omega=\Gamma$ to be $\Delta E_{j=1,\ldots,5}=\Delta E \exp\big(i\frac{2(j-1)\pi}{5}\big)$, with the same amplitude
\begin{equation}
\Delta E=\left(\frac{36\pi\Omega^4}{m}|\psi_0(0)|^2 \epsilon\right)^{\frac{1}{5}}. \label{fifth_root}
\end{equation}
The expression (\ref{fifth_root}) fits well with the exact numerical solution in Fig.\ref{fig3}(d). This is the typical feature for a fifth-order EP. Indeed, for an isotropic ($M$-independent) interaction, the five energy levels (as well as the associated eigenstates) merge at $\Gamma=\Omega$, see Fig.\ref{fig3}(c), which marks the location of a fifth-order EP.

To this end, we have demonstrated the existence of third-order and fifth-order EPs for two bosons with spin-1/2 and spin-1, and the ability to achieve $\sim \epsilon^{1/3}$ and $\sim \epsilon^{1/5}$ spectral sensitivity by introducing a small interaction anisotropy within the ferromagnetic  scattering channel ($S=1$ for spin-1/2 bosons and $S=2$ for spin-1 bosons).

\section{Three-boson system} \label{sec4}

In this section, we study the properties of three spin-1/2 and spin-1 boson systems. Here we take the mean-field treatment as established by exact two-body solutions in Sec.\ref{sec3}, which assumes the charge parts of all three bosons are frozen at the lowest harmonic oscillator level ($n=0$). We will discuss the spectral response to interaction anisotropy in both two-body and three-body coupling sectors. %, and finally discuss the mathematical origin for the obtained spectral sensitivity.

\subsection{Spin-1/2}

For three spin-1/2 bosons, the allowable total spin are $S=3/2,1/2$ respectively. The wave-function of ferromagnetic $S=3/2$ state is fully symmetric and the ground state under an s-wave interaction belongs to this spin space. Then it is naturally followed that with an isotropic($M$-independent) interaction, a fourth-order EP will be supported at $\Gamma=\Omega$, where all four energy levels and eigen-states coalesce simultaneously. In the next we will mainly focus on the mean-field spectral response to interaction anisotropies, which can be imposed on the two-body or three-body collision channel.

\begin{figure}[h]
\includegraphics[width=8.5cm]{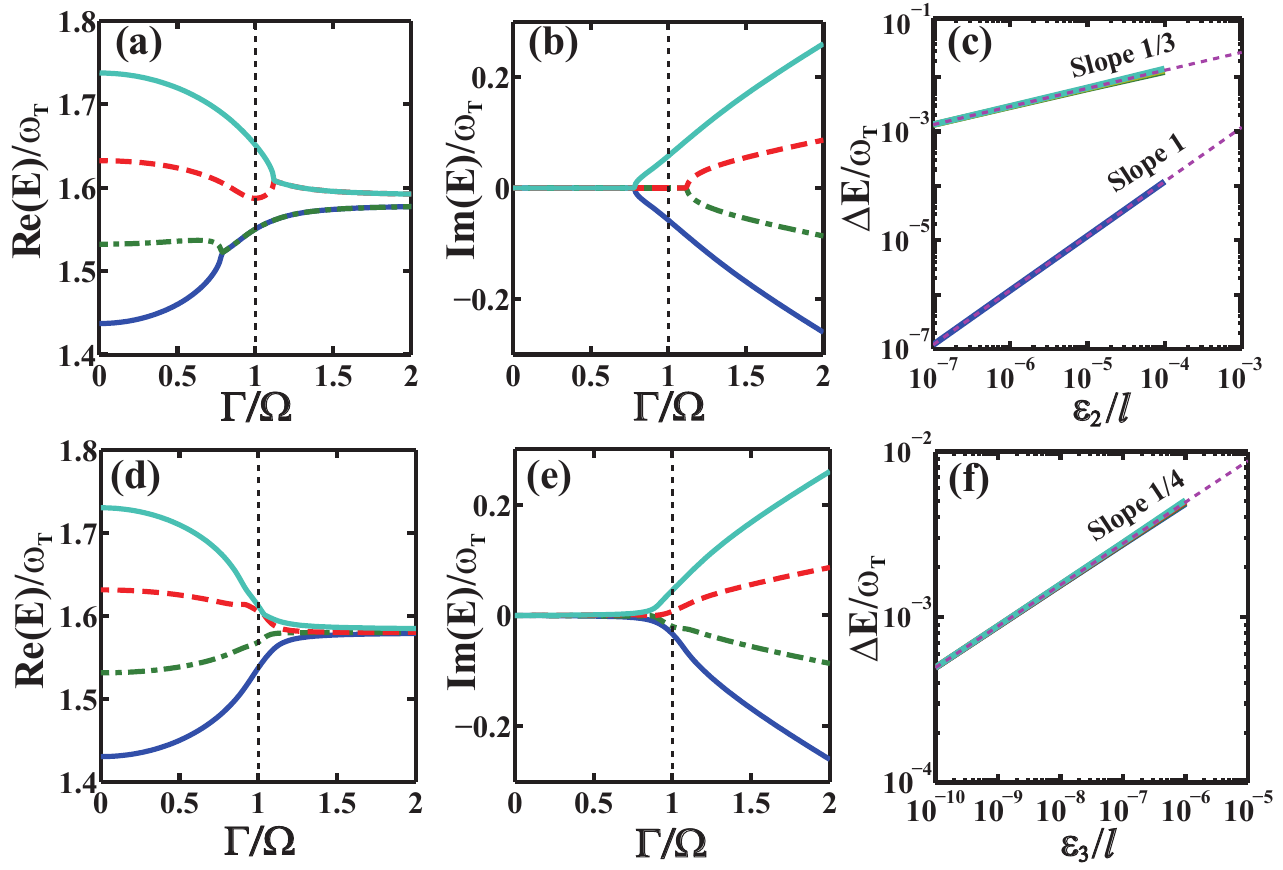}
\caption{(Color online).
Mean-field spectral responses of three harmonically trapped spin-1/2 bosons with anisotropic interactions in the two-body (a,b,c) and three-body (d,e,f) collision channels. (a,b) shows the real and imaginary parts of energy levels as a function of $\Gamma$ under a two-body interaction anisotropy:  $a_s^{1,1}=a_s^{1,-1}=a_s$ and $a_s^{1,0}=a_s+\epsilon_2$, where $a_s\equiv l/30$ and $\epsilon_2=0.005l$. (c) shows the according energy shifts of four levels at $\Gamma/\Omega=1$ as a function of $\epsilon_2$. (d,e) shows the spectral responses with an isotropic two-body interaction and an additional three-body interaction ($\epsilon_3=0.005l$) among $\uparrow\uparrow\uparrow$ sector. (f) shows the according energy shifts of four levels at $\Gamma/\Omega=1$ as a function of $\epsilon_3$. The dashed lines in (d,f) show analytical fit (see text). In all plots, we take $\Omega/\omega_T=0.1$.} \label{fig4}
%\end{minipage}
\end{figure}

For the two-body interaction anisotropy, we consider an anisotropy in $S=1,M=0$ channel (with spin state $|\uparrow\downarrow\rangle+|\downarrow\uparrow\rangle$) and plot the resulted spectrum in Fig.\ref{fig4} (a,b). Different from the isotropic case where the four levels simultaneously coalesce at EP ($\Gamma/\Omega=1$), here at this point two of the levels become complex and the rest two are still real. This means that the two eigenvectors associated with the two real eigen-energies are linearly independent. In Fig.\ref{fig4} (c), we plot the amplitudes of four energy shifts as a function of two-body anisotropy strength ($\epsilon_2$), and find that three levels obey a cube-root response $\sim\epsilon_2^\frac{1}{3}$, and the rest one shows a linear response $\sim\epsilon_2$.

The above results can be understood by writing the corresponding matrix representation of Hamiltonian in the ferromagnetic spin basis $\{|S=\frac{3}{2},M=\frac{3}{2}\rangle,|S=\frac{3}{2},M=\frac{1}{2}\rangle,|S=\frac{3}{2},M=-\frac{1}{2}\rangle,|S=\frac{3}{2},M=-\frac{3}{2}\rangle\}$. Given the two-body interaction anisotropy in $|\uparrow\downarrow\rangle+|\downarrow\uparrow\rangle$ channel, we have (up to a constant mean-field shift)
%\begin{widetext}
\begin{equation}
\begin{split}
H_{MF}(\epsilon_2)=%\frac{4\pi a_s}{m}|\psi_0(0)|^2\Bbb1+
\left( \begin{array}{ccccc}
  \frac{3}{2}i\Gamma & \frac{\sqrt{3}}{2}\Omega & 0 & 0 \\
  %\frac{\sqrt{3}}{2}\Omega  & \frac{1}{2}i\Gamma+\frac{8\pi a_s}{m}|\psi_0(0)|^2\epsilon & \Omega & 0 \\
  \frac{\sqrt{3}}{2}\Omega  & \frac{1}{2}i\Gamma+2\delta & \Omega & 0 \\
  %0  & \Omega & -\frac{1}{2}i\Gamma+\frac{8\pi a_s}{m}|\psi_0(0)|^2\epsilon & \frac{\sqrt{3}}{2}\Omega \\
  0  & \Omega & -\frac{1}{2}i\Gamma+2\delta & \frac{\sqrt{3}}{2}\Omega \\
  0  & 0 & \frac{\sqrt{3}}{2}\Omega & -\frac{3}{2}i\Gamma \\
\end{array}
\right)
\label{H_3-half}
\end{split}
\end{equation}
%\end{widetext}
with $\delta=\frac{4\pi a_s}{m}|\psi_0(0)|^2\epsilon_2$. Then it is straightforward to show that at EP and under a small $\epsilon$($\ll l$),  one energy splitting shows a linear response $\sim \frac{6\pi}{m}|\psi_0(0)|^2\epsilon_2$, and the rest three are $\Delta E_{j=1,2,3}=\Delta E\exp\big(i\frac{2(j-1)\pi}{3}\big)$, with the amplitude $\Delta E$ following a cube-root relation:
\begin{equation}
\Delta E=\left(\frac{24\pi\Omega^2}{m}|\psi_0(0)|^2 \epsilon_2\right)^{\frac{1}{3}}.
\end{equation}
This can well explain the results shown in Fig.\ref{fig4}(c).

In order to fully break the energy degeneracy at the fourth-order EP and create the most sensitive spectral response $\epsilon^{1/4}$, we introduce a more sophisticated perturbation to the system, i.e., an interaction anisotropy in the three-body collision sector. Here we choose a three-body anisotropic interaction($\epsilon_3$) in  $|\uparrow\uparrow\uparrow\rangle$ channel for example, and shows the resulted spectrum in Fig.\ref{fig4}(d,e).  One can see that at EP ($\Gamma=\Omega$), the four-fold degeneracy are fully broken and all the four energy levels develop imaginary parts. Fig.\ref{fig4}(f) shows that the energy shifts indeed obeys the $\sim \epsilon_3^{1/4}$ dependence. Similarly, these results can be conveniently understood by writing down the Hamiltonian in ferromagnetic basis, which reads
\begin{equation}
\begin{split}
H_{MF}(\epsilon_3)=%\frac{4\pi a_s}{m}|\psi_0(0)|^2\Bbb1+
\left( \begin{array}{ccccc}
  \frac{3}{2}i\Gamma+\frac{4\pi a_s}{m}|\psi_0(0)|^2\epsilon_3 & \frac{\sqrt{3}}{2}\Omega & 0 & 0 \\
  \frac{\sqrt{3}}{2}\Omega  & \frac{1}{2}i\Gamma & \Omega & 0 \\
  0  & \Omega & -\frac{1}{2}i\Gamma & \frac{\sqrt{3}}{2}\Omega \\
  0  & 0 & \frac{\sqrt{3}}{2}\Omega & -\frac{3}{2}i\Gamma \\
\end{array}
\right)
\label{H_3-half2}
\end{split}
\end{equation}
This gives a fully fourth-root energy splitting $\Delta E_{j=1,2,3,4}=\Delta E\exp\big(i\frac{(4j-3)\pi}{8}\big)$ with
\begin{equation}
\Delta E= \left(\frac{3\pi\Omega^3}{m}|\psi_0(0)|^2\epsilon_3\right)^{\frac{1}{4}}
\end{equation}
for small $\epsilon_3$. The numerical results in Fig.\ref{fig4}(f) verifies this spectral response.
%{\color{red} Take interaction anisotropy only in up-up-up channel but not including down-down-down channel}

\subsection{Spin-1}

In this subsection, we study three spin-1 bosons whose allowable total spin values are $S=3,2,1,0$. Like in the two-body case discussed previously, we focus on the case of ground state with ferromagnetic order, such as in $^{87}$Rb atoms with $a_2<a_0$. In this case, the ground state of three bosons lies in the total spin $S=3$ sector, which include seven magnetic states. Given an $M$-independent interaction, a seventh-order EP can be supported at $\Gamma=\Omega$, where all seven energy levels and eigenstates coalesce simultaneously.

\begin{figure}[h]
\includegraphics[width=8.5cm]{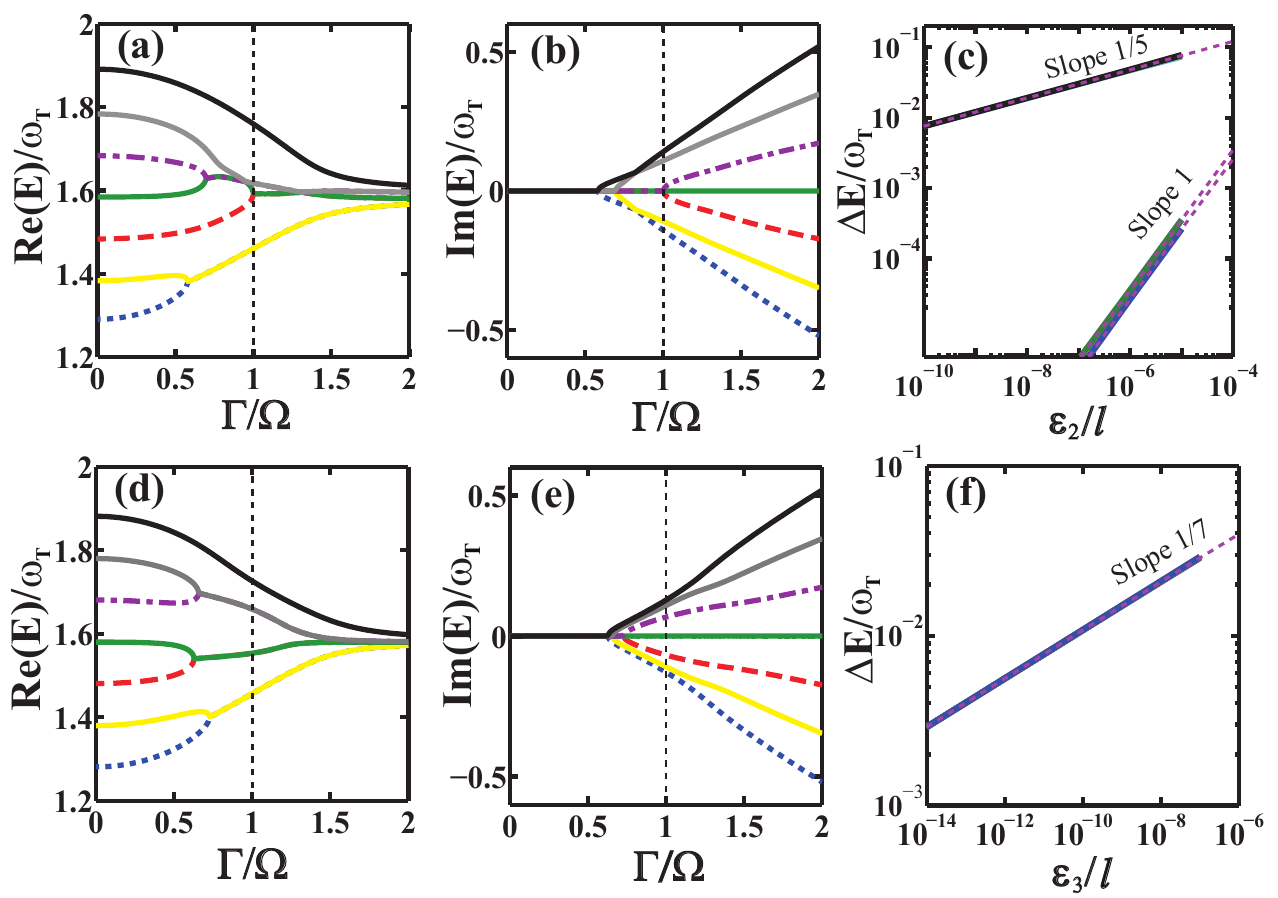}
\caption{(Color online).
Mean-field spectral responses of three harmonically trapped spin-1 bosons with anisotropic interactions in the two-body (a,b,c) and three-body (d,e,f) collision channels. (a,b) shows the   real and imaginary parts of energy levels as a function of $\Gamma$ under a two-body interaction anisotropy in $S=2,M=0$ channel:  $a_s^{2,\pm2}=a_s^{2,\pm 1}=a_s$ and $a_s^{2,0}=a_s+\epsilon_2$, where $a_s= l/30$ and $\epsilon_2=0.005l$. (c) shows the according energy shifts of seven levels at $\Gamma=\Omega$ as a function of $\epsilon_2$. (c,d) shows the spectral responses with an isotropic two-body interaction and an additional three-body interaction ($\epsilon_3=0.005l$) in $S=3,M=0$ channel. (f) shows the according energy shifts of seven levels at $\Gamma=\Omega$ as a function of $\epsilon_3$. The dashed lines in (d,f) shows analytical fit (see text). In all plots, we take $\Omega/\omega_T=0.1$.} \label{fig5}
\end{figure}

Now we turn to the spectral response under interaction anisotropies. For two-body interaction anisotropy, we consider the same one as in Sec.\ref{spin1}, i.e., $a_s^{S=2,M=\pm 2,\pm 1}=a_s$ and $a_s^{2,0}=a_s+\epsilon_2$. The resulted spectrum for the lowest seven levels is shown in Fig.\ref{fig5}(a,b) for typical $a_s$ and $\epsilon_2$. In this case, five levels obey a sensitive response $\sim \epsilon_2^{1/5}$, and the rest two give linear responses $\sim \epsilon_2$, see Fig.\ref{fig5}(c). To explain this, we write down the corresponding matrix representation of Hamiltonian in the ferromagnetic basis $\{|S=3,M\rangle\}$ (integer $M\in[-3,3]$) as %\begin{widetext}
\begin{equation}
\begin{split}
%H_{MF}(\epsilon_2)=%\frac{4\pi a_s}{m}|\psi_0(0)|^2\Bbb1+
\left( \begin{array}{ccccccc}
  3i\Gamma & \frac{\sqrt{6}}{2}\Omega & 0 & 0 & 0 & 0 & 0\\
  \frac{\sqrt{6}}{2}\Omega & 2i\Gamma & \frac{\sqrt{10}}{2}\Omega & 0 & 0 & 0 & 0 \\
  0 & \frac{\sqrt{10}}{2}\Omega & i\Gamma+\frac{12\delta}{5} & \sqrt{3}\Omega  & 0 & 0 & 0\\
  0 & 0 & \sqrt{3}\Omega & \frac{18\delta}{5} & \sqrt{3}\Omega & 0 & 0\\
  0 & 0 & 0 & \sqrt{3}\Omega & -i\Gamma+\frac{12\delta}{5} & \frac{\sqrt{10}}{2}\Omega & 0\\
  0 & 0 & 0 & 0 & \frac{\sqrt{10}}{2}\Omega & -2i\Gamma & \frac{\sqrt{6}}{2}\Omega\\
  0 & 0 & 0 & 0 & 0 & \frac{\sqrt{6}}{2}\Omega & -3i\Gamma
\end{array}
\right)
\label{H_5+1+1}
\end{split}
\end{equation}
%\end{widetext}
here $\delta\equiv \frac{4\pi}{m}|\psi_0(0)|^2\epsilon_2$. In order to obtain the energy shifts analytically, we write out the secular equation for eigenvalues
%\begin{widetext}
\begin{equation}
\begin{split}
-E^7&+\frac{6}{5}\delta(7E^6+36E^4\Omega^2+165E^2\Omega^4)\\
&+\frac{36}{25}\delta^2(16E^5+130 E^3\Omega^2 +675E\Omega^4)\\
&+\frac{648}{125}\delta^3(4E^4+40 E^2\Omega^2+225\Omega^4)=0
\end{split}
\end{equation}
%\end{widetext}
For small $\delta\propto \epsilon_2$, we can extract two independent equations by comparing their order:
%when $\epsilon\rightarrow0$
\begin{equation}
\begin{split}
-E^5+78\Omega^4\delta=0; 55E^2\delta-270\delta E+324\delta^2=0
\end{split}
\end{equation}
which give two linear solutions $\sim (\frac{27}{11}\pm\frac{9\sqrt{5}}{55})\delta$ and five solutions following a fifth-root relation
$\Delta E_{j=1,\ldots,5}=\Delta E\exp\big(i\frac{2(j-1)\pi}{5}\big)$, where
\begin{equation}
\Delta E=\left( \frac{312\pi\Omega^4}{m}|\psi_0(0)|^2\epsilon_2 \right)^{\frac{1}{5}}.
\end{equation}
The corresponding numerical results in Fig.\ref{fig5}(c) confirm this conclusion.

To generate a more sensitive spectral response, similar to the spin-1/2 case, we turn on an interaction anisotropy in the three-body collision channel. For simplicity, we consider a three-body anisotropy ($\epsilon_3$) for three bosons colliding in $S=3,M=0$ channel, and the resulted spectrum are shown in Fig.\ref{fig5}(d,e). We can see that the splitting of the spectrum shows a distinct structure as compared to the two-body anisotropy case in Fig.\ref{fig5}(a,b). In this case the amplitudes of energy shifts at EP identically scales as $\sim \epsilon_3^{1/7}$, as shown in Fig.\ref{fig5}(f). This is the most sensitive response of a seventh-order EP to small perturbations. In this case, the corresponding Hamiltonian in the ferromagnetic basis is
%\begin{widetext}
\begin{equation}
\begin{split}
%H_{MF}^1(\epsilon_3)=%\frac{4\pi a_s}{m}|\psi_0(0)|^2\Bbb1+
\left( \begin{array}{ccccccc}
  3i\Gamma & \frac{\sqrt{6}}{2}\Omega & 0 & 0 & 0 & 0 & 0\\
  \frac{\sqrt{6}}{2}\Omega & 2i\Gamma & \frac{\sqrt{10}}{2}\Omega & 0 & 0 & 0 & 0 \\
  0 & \frac{\sqrt{10}}{2}\Omega & i\Gamma & \sqrt{3}\Omega  & 0 & 0 & 0\\
  0 & 0 & \sqrt{3}\Omega & \frac{4\pi}{m}|\psi_0(0)|^2\epsilon_3 & \sqrt{3}\Omega & 0 & 0\\
  0 & 0 & 0 & \sqrt{3}\Omega & -i\Gamma & \frac{\sqrt{10}}{2}\Omega & 0\\
  0 & 0 & 0 & 0 & \frac{\sqrt{10}}{2}\Omega & -2i\Gamma & \frac{\sqrt{6}}{2}\Omega\\
  0 & 0 & 0 & 0 & 0 & \frac{\sqrt{6}}{2}\Omega & -3i\Gamma
\end{array}
\right)
\label{H_7}
\end{split}
\end{equation}
%\end{widetext}
For small anisotropy $\epsilon_3$, the energy splitting of the seven levels at EP are $\Delta E_{j=1,\ldots,7}=\Delta E\exp{\big(i\frac{2(j-1)}{7}}\big)$ with the same amplitude
\begin{equation}
\Delta E=\left(\frac{900\pi\Omega^6}{m}|\psi_0(0)|^2\epsilon_3\right)^{\frac{1}{7}}
\end{equation}
This analytical result is consistent with the numerical results shown in Fig.\ref{fig5}(f).

\section{Mathematical origin for the spectral sensitivity and generalization to many-body systems}  \label{sec5}

In the previous section, we have shown the spectral response for small cluster boson systems to different types of interaction anisotrpies. In Table \ref{table1}, we summarize the spectral response in terms of the interaction anisotropy strength $\epsilon$ for spin-1/2 and spin-1 boson systems with particle number $N$ and under $k$-body interaction anisotropy.

\begin{table}[h]
\begin{ruledtabular}
\begin{tabular}{c|cc}
 &   $s=\frac{1}{2}$ & $s=1$ \\ \hline
$N=2$, $k=2$ & $\epsilon^{1/3}$ &  $\epsilon^{1/5}$		\\
$N=3$, $k=2$ & $\epsilon^{1/3}$ &  $\epsilon^{1/5}$		\\
$N=3$, $k=3$ & $\epsilon^{1/4}$ &  $\epsilon^{1/7}$		\\
\end{tabular}
\end{ruledtabular}
\caption{\label{table1}Spectral responses of spin-$s$ boson systems with particle number $N$ and under $k$-body interaction anisotropy with strength $\epsilon$.  }
\end{table}

%One could ask what's the origin behind these response properties. We could come up with that they should be relevant to the matrix structure of perturbation.
We emphasis that results in Table.\ref{table1} universally depend on $N$, $k$ and spin $s$, but not on the concrete form of perturbation, i.e., the specific channel of interaction anisotropy. This implies there exist a robust intrinsic mechanism for the phenomenon. In our previous work\cite{Pan}, we have unveiled such mechanism for spin-1/2 bosons. Here we will illustrate the idea for small cluster spin-1 bosons, and finally extend to systems with an arbitrary $N,\ k$ and $s$.

As we mainly focus on the spectral sensitivity of the EP system and the perturbation is induced by an anisotropy in spin channel, in later discussions we only consider the spin-dependent Hamiltonian at EP ($\Gamma=\Omega$):
\begin{equation}
H_{sd}=\Omega (S_x+i S_z)+H';  \label{Ham_tot}
\end{equation}
here $H'$ refers to the perturbation part induced by interaction anisotropies.

\subsection{Spectral sensitivity for small cluster spin-1 bosons}

We will discuss three cases listed in Table \ref{table1} for spin-1 bosons.

{\bf (I) Two bosons with two-body anisotropy ($N=2,\ k=2$):}

For the two-body ground state in $S=2$ subspace, the Hamiltonian (\ref{Ham_tot}) is expanded by a $5\times5$ matrix. In the absence of $H'$, $H_{sd}$ results in a fifth-order EP, which can be understood conveniently by a spin rotation around x-axis. Specifically, under a unitary transformation
\begin{equation}
U=e^{i\frac{\pi}{2}S_x}, \label{U}
\end{equation}
we have $UH_{sd}U^{-1}=\Omega S_+$. This shows that $H_{sd}$ simply produces the angular momentum raising operator, which has only one eigenstate $|S=2,M=2\rangle=|1,0,0,0,0\rangle$ with eigenvalue $0$. On the other hand, the raising operator $S_+$ is associated with an fifth-order Jordan block in the spectral decomposition. Both properties justify the occurrence of a fifth-order EP in this five dimensional spin space.

Next, consider the perturbation in $|S=2,M=0\rangle$ channel as discussed in Sec.\ref{spin1}, which is proportional to the spin projection operator  $P_{S=2,M=0}\sim (S_z-2)(S_z-1)(S_z+1)(S_z+2)$, here $S_z$ is the z-component of total spin operator. Under the same spin rotation $U$, this perturbation can be expressed as $(S_y-2)(S_y-1)(S_y+1)(S_y+2)$, giving the following matrix in the five dimensional spin space
\begin{equation}
H'_{rotated}\left(
  \begin{array}{ccccc}
    * & 0 & * & 0 & *\\
    0 & * & 0 & * & 0\\
    * & 0 & * & 0 & *\\
    0 & * & 0 & * & 0\\
    * & 0 & * & 0 & *\\
  \end{array}
\right)_{5\times 5}
\label{2spin-1_2b_1}
\end{equation}
where $*$ refers to the non-zero element proportional to the perturbation parameter $\epsilon$. By straightforward algebra, we can see that this type of perturbation, together with the $S_+$ operator, can give rise to an eigenvalue (equal to the energy splitting at EP) as $\epsilon^{1/5}$. More importantly, this analysis allows us to extend to other types of two-body perturbation. In general, if one considers the perturbation in $S_z=M$ scattering channel,
\begin{equation}
H'\sim \prod_{m\neq M}(S_z-m); \label{general}
\end{equation}
then under rotation $U$, $H'$ becomes $ \prod_{m\neq M}(S_y-m)$, which possesses the following type of matrix
\begin{equation}
H'_{rotated}=\left(
  \begin{array}{ccccc}
    * & * & * & * & *\\
    * & * & * & * & *\\
    * & * & * & * & *\\
    * & * & * & * & *\\
    * & * & * & * & *\\
  \end{array}
\right)_{5\times 5}
\label{2spin-1_2b_2}
\end{equation}
This perturbation matrix can still provide an energy splitting $\epsilon^{1/5}$. This is because the two matrices (\ref{2spin-1_2b_1}) and (\ref{2spin-1_2b_2}) both include non-zero elements generated by $S_{\pm}^4$. Therefore the qualitative form of energy splitting $\sim\epsilon^{1/5}$ still holds, which does not depends on the specific spin channel of the anisotropic interaction.

{\bf (II) Three bosons with two-body anisotropy ($N=3,\ k=2$):}

Different from the case (I) where the ferromagnetic state of two bosons is with total spin $S=2$, here for three bosons the ferromagnetic state is with spin $S=3$. Thus the dimension of Hamiltonian matrix is enlarged to seven. In this case, we write down a general type of two-body interaction anisotropy:
\begin{equation}
H'\sim \sum_{\langle i,j\rangle} \prod_{m\neq M}(S^{ij}_z-m); \label{general}
\end{equation}
here $S^{ij}_z=s_{z,i}+s_{z,j}$.  After the rotation $U$ (\ref{U}), we get the following matrix structure for the perturbation Hamiltonian
\begin{equation}
H'_{rotated}=\left(\begin{array}{ccccccccc}
* & * & * & * & * & 0 & 0\\
    * & * & * & * & * & *& 0\\
    * & * & * & * & * & * & *\\
    * & * & * & * & * & * & *\\
    * & * & * & * & * & * & *\\
    0 & * & * & * & * & * & *\\
    0 & 0 & * & * & * & * & *\\
\end{array}\right)_{7\times 7}
\label{3spin-1_2b_2}
\end{equation}
We can see that in comparison to case (I), in the present case  although the dimension of matrix is enlarged to seven, the non-zero elements in the matrix still extend to the fourth super- and sub-diagonals at most, because of the same form of perturbation Hamiltonian (\ref{general}). This leads to the same sensitivity for the spectral response, or energy splitting, as $\epsilon^{1/5}$.

{\bf (III) Three bosons with three-body anisotropy ($N=3,\ k=3$):}

Different from case (II), in the present case the perturbation is in three-body collision channel. For the ferromagnetic state of three bosons ($S=3$), such perturbation Hamiltonian can be written as certain superposition of projection operators in $S=3$ spin space:
\begin{eqnarray}
H' &=& \sum_M c_M P_{S=3,M} \nonumber\\
&\sim& \sum_M c_M \prod_{m\neq M}(S_z-m); \label{general_3}
\end{eqnarray}
here each $P_{S=3,M}$ is a six-rank polynomial of $S_z$; for instance, we have $P_{S=3,M=3}\sim (S_z-2)(S_z-1)S_z(S_z+1)(S_z+2)(S_z+3)$, with $S_z$ the z-component of total spin operator for three bosons. After the spin rotation $U$(\ref{U}), $H'$ becomes a six-rank polynomial of $S_y$ (or equivalently $S_{\pm}$), which leads to the following matrix structure:
\begin{equation}
H'_{rotated}=\left(\begin{array}{ccccccccc}
* & * & * & * & * & * & *\\
    * & * & * & * & * & *& *\\
    * & * & * & * & * & * & *\\
    * & * & * & * & * & * & *\\
    * & * & * & * & * & * & *\\
    * & * & * & * & * & * & *\\
    * & * & * & * & * & * & *\\
\end{array}\right)_{7\times 7}
\label{3Spin-1_3b}
\end{equation}
Here the non-zero elements can extend to the top right and the lower left corners of $7\times 7$ matrix, thus giving rise to $\epsilon^{1/7}$ spectral response. This is the conclusion that is irrelevant to the specific channel for the three-body anisotropy.

%Moreover, the spectral response is always $\epsilon^{\frac{1}{7}}$ in any other spin channel as long as the anisotropic interaction is induced by three-body interaction.\\

\subsection{Generalization to many-body system with an arbitrary spin}

From the previous subsection, we know that the order of EP and the spectral sensitivity at EP universally depend on a few parameters, namely, the particle number $N$, the spin $s$, and the number of colliding particles for the anisotropic interaction $k$ (for instance, $k=2$ means the interaction anisotropy in two-body coupling sector).

First, for spin-$s$ bosons with particle number $N$, if the ground state is ferromagnetic with total spin $S=Ns$, the EP order can be as high as  $2Ns+1$. In this case, given a spin-independent interaction, the spin-dependent part of Hamiltonian is solely give by the $\mathcal{PT}$ potential $H_{PT}=\Omega (S_x+i S_z)$, where $S_{\alpha}=\sum_j s_{j,\alpha}$ is the total spin operator in $\alpha=x,y,z$ component. The many-body bosons in the ferromagnetic state just behaves as a huge spin with $S=Ns$. Then following the same analysis in previous sections, upon a spin rotation around x (see Eq.\ref{U}), $H_{PT}$ simply reproduces the angular momentum raising operator $S_+$, which has one and only one eigenstate $|S_z=Ns\rangle=|1,0,...0\rangle$.  Moreover, the raising operator is associated with an ($2Ns+1$)th-order Jordan block in the spectral decomposition. These properties justify the occurrence of ($2Ns+1$)th-order EPs in ($2Ns+1$)-dimensional spin space.

\begin{figure}[h]
\includegraphics[width=8.5cm]{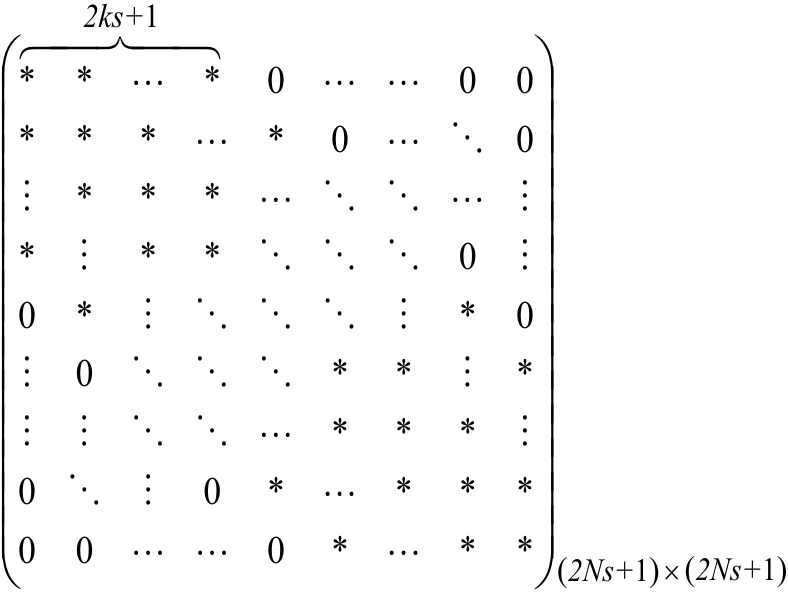}
\caption{(Color online).
$(2ks+1)$-Hessenberg matrix produced by spin-s bosons with number $N$ and $k$-body interaction anisotropy. The matrix has only zero entries below/above the $(2ks+1)$-th sub/super diagonals (including the main diagonal).\\
%{\color{red}Change "k" to $2ks+1$, and "N" to $2Ns+1$ in the figure}
} \label{fig6}
\end{figure}

Second, given the high EP order $2Ns+1$, the spectral sensitivity induced by the interaction anisotropy will additionally depend on $k$. Namely, the  $k$-body interaction anisotropy can be described by projection operators on spin-$ks$, which gives rise to $H'$ as a $2ks$-rank polynomial of $S_z$. After a rotation $U$,  $H'$ becomes  $2ks$-rank polynomial of $S_y$ (or equivalently $S_{\pm}$). This leads to the matrix structure as shown in Fig.\ref{fig6}, where the non-zero elements can extend to the $(2ks+1)$-th super- and sub-diagonals (including the main diagonal).

The structure of matrix shown in Fig.\ref{fig6} is exactly the $(2ks+1)$-Hessenberg matrix constituted by Jordan blocks $J_{2Ns+1}$ under perturbations\cite{Ma}.  Mathematically, upper(lower) $q$-Hessenberg matrix is a matrix with only zero entries below(above) the $q$-th (including the main diagonal) sub-diagonal (super-diagonal). This kind of  matrix can lead to $[\frac{2Ns+1}{2ks+1}]$ groups of sub-EPs, and each gives rise to the spectral splitting as $\epsilon^{\frac{1}{2ks+1}}$ at best.

From above results, we remark that although the EP order depends on the particle number $N$, the actual spectral sensitivity at EP does not rely on $N$, but solely depends on $k$ and $s$. For instance, for case (I) and (II) in previous subsection, although the particle number $N$ (and thus the EP order) are different, the spectral sensitivities are the same for (I) and (II) (both are $\epsilon^{1/5}$) because $k,s$ do not change. Therefore, to create the spectral response as sensitive as possible, one has to resort to higher spin ($s$ large) and higher-body collision channel ($k$ large) but not to more particles ($N$ large).

In principle, above results for bosons can be extended to fermionic systems. However, the full ferromagnetic state is usually not favored by fermion statistics, which makes the analysis of fermions not as transparent as bosons. For instance, for two spin-1/2 fermions,
the scattering is in singlet channel, while the $\mathcal{PT}$ potential takes no effect to singlet state because $H_{PT}|S=0\rangle=0$, similar to spin-1 bosons in the anti-ferromagnetic channel. For two spin-3/2 fermions, the allowable total spin is $S=0,2$; the only scattering channel that the $\mathcal{PT}$ potential can take effect is $S=2$ channel, which will leads to a fifth-order EP, similar to spin-1 bosons in the ferromagnetic channel.  The extension to many fermions and to arbitrary spin can be similarly analyzed, which will not be elaborated here.

\section{Experimental detection}  \label{sec6}

In the previous sections we have shown the ultra-sensitive spectral response of high-order EPs in the presence of small perturbations. Such response could be detected by the radio-frequency (rf) spectroscopy, which has been widely used in cold atoms experiments. In the following, we propose another scheme to detect these spectral responses, i.e., through the dynamical approach.

To resolve the quantum dynamics of a non-Hermitian system, the conventional way is to use the  Lindblad master equation, which results from the Markovian approximation\cite{Breuer} of the reservoir and describes the dissipative evolution of the density matrix $\varrho$:
%Note that the non-Hermitian Hamiltonian is always restricted to a phenomenological description.
\begin{eqnarray}
\frac{d\varrho}{dt}&=&-i[H_s,\varrho]+\sum_k\Big(L_k\varrho L_k^{\dag}-\frac{1}{2}\{L_k^{\dag}L_k,\varrho\}\Big)\nonumber \\
&=&-i\big(\mathcal{H}\varrho-\varrho \mathcal{H}^{\dag}\big)+\sum_kL_k\varrho L_k^{\dag}. \label{LME1}
\end{eqnarray}
Here $H_s$ is the Hermitian Hamiltonian of the system in the absence of losses; $L_k$'s are the Lindblad dissipation operators;  $\mathcal{H}=H_s-\frac{i}{2}\sum_kL_k^{\dag}L_k$ is the non-Hermitian Hamiltonian with losses. %and for the two-body problem it is exactly the model Hamiltonian (\ref{Ham_2b}) as we have used. By equating $H_{eff}$ and $H$ in (\ref{Ham_2b}), one can determine two sets of  Lindblad operators as: $L_{1}=..,\ L_{1}^{\dag}=$ and $L_{1}=..,\ L_{1}^{\dag}=..$.
%, in which case the Lindblad master equation will have the following form
%\begin{eqnarray}
%\frac{d\rho}{dt}&=&-i[H_s,\rho]-\frac{\Gamma}{2}\{b_\downarrow^{\dag}b_\downarrow,\rho\}
%+\frac{\Gamma}{2}\{b_\uparrow^{\dag}b_\uparrow,\rho\} \nonumber \\
%&+&\Gamma b_\downarrow\rho b_\downarrow^{\dag}-\Gamma b_\uparrow\rho b_\uparrow^{\dag}
%\label{LME2}
%\end{eqnarray} in (\ref{Ham_2b})
Consider the case illustrated in Fig.\ref{fig1}(a), we choose the
%The $\mathcal{PT}$-symmetric Hamiltonian with balanced gain and loss can be expressed by appropriate choice of Lindblad operators. Here we choose the
Lindblad operator $L_k^{\dag}=\sqrt{2\Gamma}a_{\downarrow}^{\dag}$, $L_k=\sqrt{2\Gamma}a_{\downarrow}$ to describe single-particle loss in spin-$\downarrow$.  In this case, we have $\mathcal{H}=H_s-i\Gamma a_{\downarrow}^{\dag}a_{\downarrow}=H_s+i\Gamma s_z-\frac{i}{2}\Gamma N$ with $N=a_{\uparrow}^{\dag}a_{\uparrow}+a_{\downarrow}^{\dag}a_{\downarrow}$. Then we can define a $\mathcal{PT}$-symmetric effective Hamiltonian as
\begin{equation}
H_{eff}\equiv\mathcal{H}+\frac{i}{2}\Gamma N=H_s+i\Gamma s_z.
\end{equation}

In order to experimentally detect the dynamical signatures of $\mathcal{PT}$-symmetric and $\mathcal{PT}$-symmetry broken phases governed by $H_{eff}$, we redefine the density matrix
\begin{eqnarray}
\rho(t)\equiv e^{N\Gamma t}\varrho(t), \label{density_matrix}
\end{eqnarray}
where the exponential factor $e^{N\Gamma t}$ is introduced to offset the pure loss term $-\frac{i}{2}\Gamma N$. The dynamics of the newly defined density matrix $\rho(t)$ can be determined by the effective Hamiltonian $H_{eff}$.

What needs to be emphasized is that by adopting the effective Hamiltonian $H_{eff}$ to describe the system, we have neglected the terms $2\Gamma a_{\downarrow}\varrho a_{\downarrow}^{\dag}$ in the Lindblad equation, which can induce the quantum jump between the diagonal density matrices in different particle number sectors\cite{Ueda,Yu}. We will show that this term does not affect the dynamics as discussed below, as long as physical quantities we studied are restricted to the initial subspace with fixed particle number. In other words, starting with $N$ atoms, the dynamics governed by $H_{eff}$ and by the Lindblad equation are consistent with each other in this $N$-body subspace.  Here, we will simply utilize the two-body spin-1/2 bosons system to illustrate our proposal, and for the sake of convenience, we still focus on the mean-field limit as discussed in previous sections.

We consider the initial state (at time $t=0$) prepared as two spin-$\uparrow$ bosons at the lowest harmonic level.
In the language of density matrix $\rho(t)$, we have $\rho_{\uparrow\uparrow}(0)\equiv\langle\uparrow\uparrow|\rho(t=0)|\uparrow\uparrow\rangle=1$ and all other matrix elements equal to zero. Under the assumption that the interaction is weak and will not induce the transition of charge part to higher harmonic levels, we have used two approaches to obtain the dynamics of $\rho_{\uparrow\uparrow}(t)$. One is to simulate the Lindblad equation (\ref{LME1}) by choosing the spin basis $\big\{|\uparrow\uparrow\rangle,|\uparrow\downarrow\rangle,|\downarrow\downarrow\rangle,
|\uparrow\rangle,|\downarrow\rangle,|0\rangle\big\}$. The other is to simulate the time-dependent Schrodinger equation under the effective Hamiltonian $H_{eff}$, which gives rise to
\begin{equation}
\rho_{\uparrow\uparrow}(t)=\langle\uparrow\uparrow|U(t)\rho(t=0)U^\dag(t)|\uparrow\uparrow\rangle  \label{rho}
\end{equation}
with $U(t)=e^{-iH_{eff}t}$. Note that the particle number is not conserved in the first approach, but conserved in the second approach. However, we can find $d\rho(t)/dt=-i\big(H_{eff}\rho-\rho H_{eff}^{\dag}\big)+2\Gamma e^{N\Gamma t}a_{\downarrow}\varrho a_{\downarrow}^{\dag}$ from the first approach which means that $d\rho_{\uparrow\uparrow}(t)/dt=-i\big(H_{eff}\rho-\rho H_{eff}^{\dag}\big)+2\Gamma e^{2\Gamma t}\langle\uparrow\uparrow|a_{\downarrow}\varrho a_{\downarrow}^{\dag}|\uparrow\uparrow\rangle$. In the initial condition $\rho_{\uparrow\uparrow}(0)=1$, the quantum jump term has no effects on $\rho_{\uparrow\uparrow}(t)$ since $\langle\uparrow\uparrow|a_{\downarrow}\varrho a_{\downarrow}^{\dag}|\uparrow\uparrow\rangle$ only relates to three-atoms term $\rho_{\downarrow\uparrow\uparrow}(0)$ which is always zero. In other words, the dynamics of $\rho_{\uparrow\uparrow}(t)$ will not be affected by the quantum jump term in the Lindblad equation in which case both approaches produce the same $\rho_{\uparrow\uparrow}(t)$.

\begin{figure}[t]
\includegraphics[width=8.5cm]{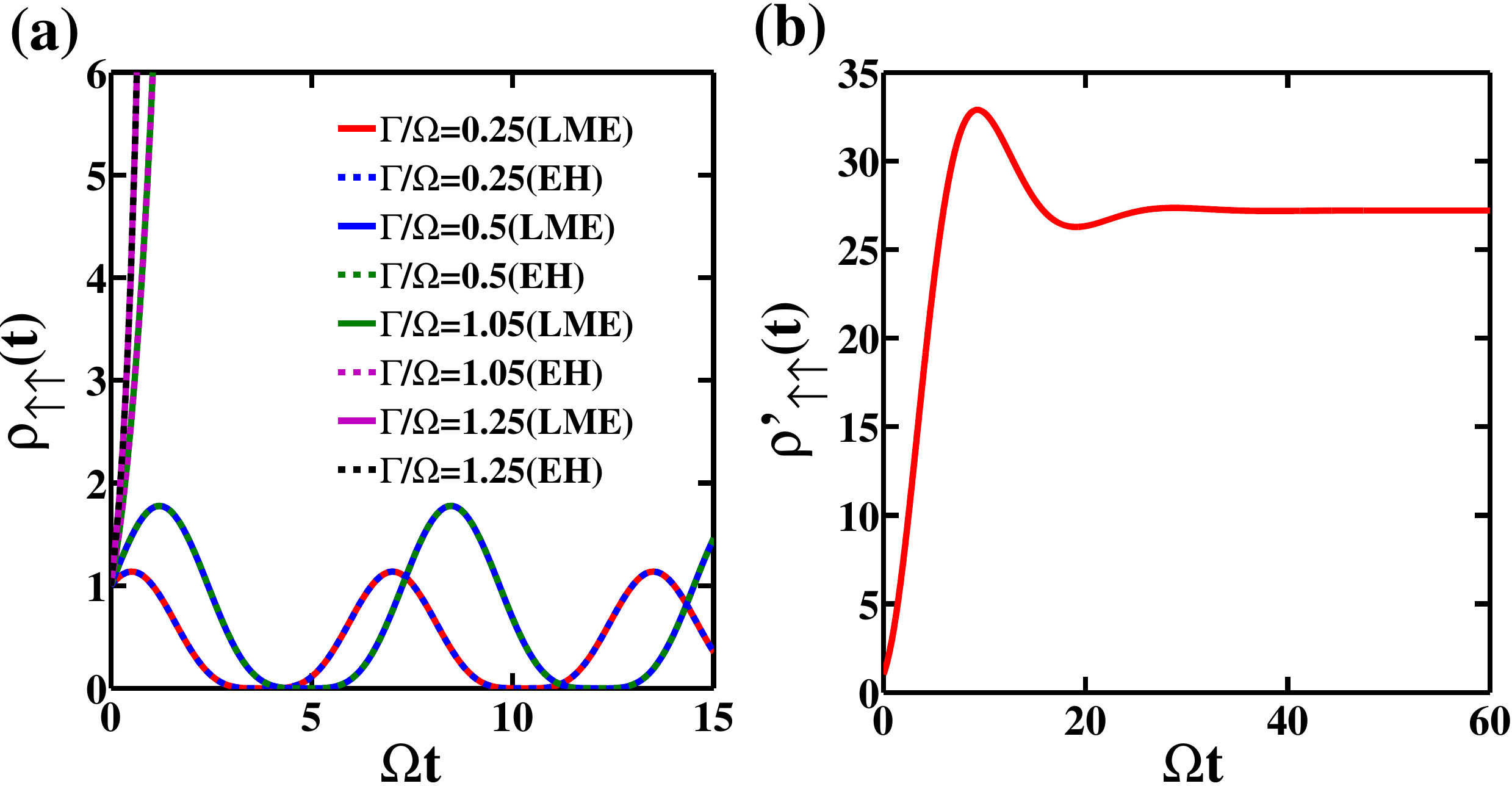}
\includegraphics[width=8.5cm]{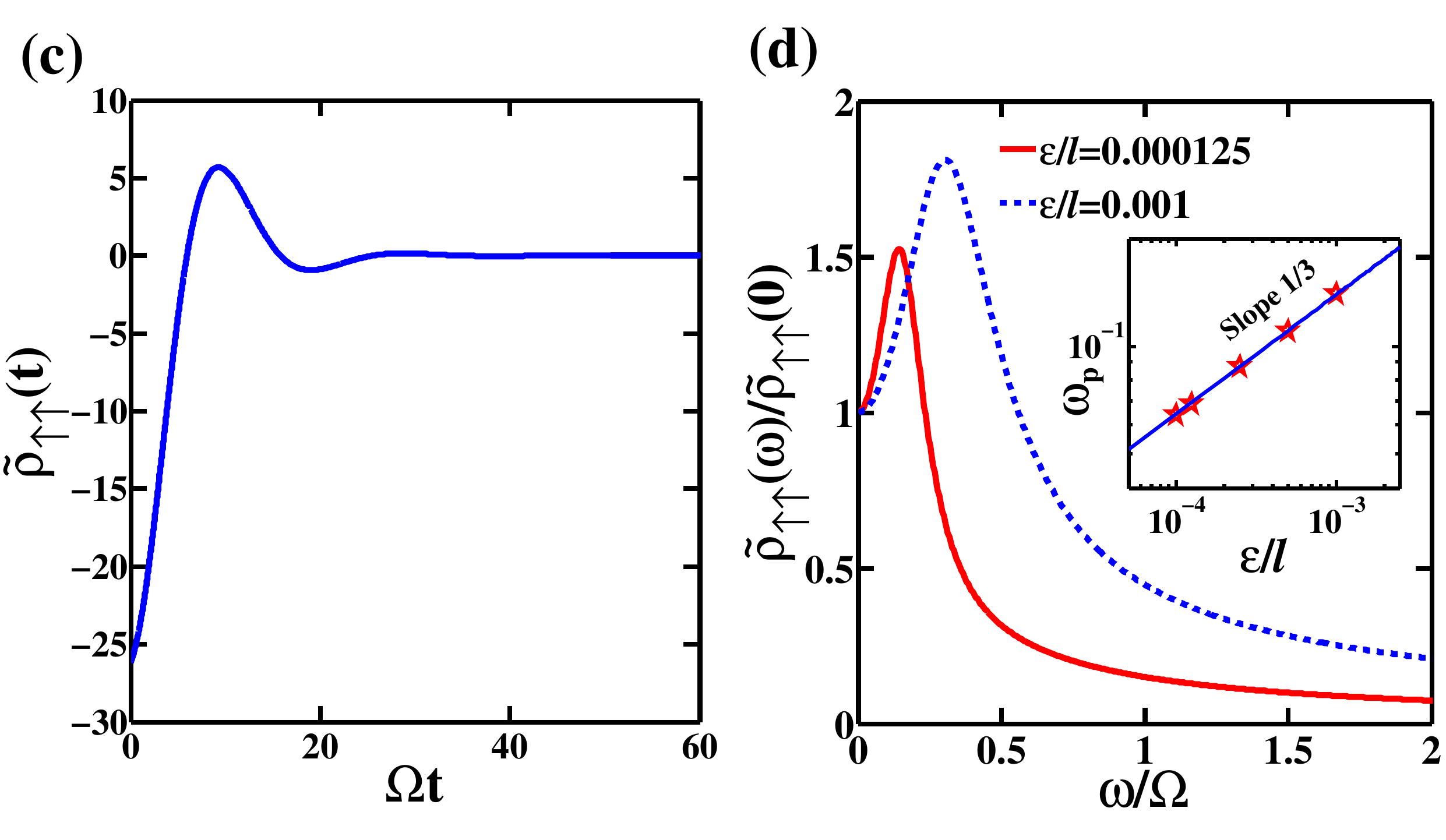}
\caption{(Color online).
(a)Dynamical evolution of  $\rho_{\uparrow\uparrow}(t)$ with initial state $\rho_{\uparrow\uparrow}(0)=1$. The oscillating curves are in the $\mathcal{PT}$-symmetric phase $\Gamma/\Omega<1$, and the monotonically increasing curves are in the $\mathcal{PT}$-symmetry broken phase $\Gamma/\Omega>1$. The solid lines are from the Lindblad master equations (LME) and the dashed lines are from the effective Hamiltonian (EH).
(b)Time evolution of $\rho'_{\uparrow\uparrow}(t)$ at EP with a tiny interaction anisotropy $\epsilon/l=0.001$.
(c)Time evolution of $\widetilde{\rho}_{\uparrow\uparrow}(t)$ at EP with a tiny interaction anisotropy $\epsilon/l=0.001$.
(d)Rescaled spectral distributions $\widetilde{\rho}_{\uparrow\uparrow}(\omega)$ at EP with tiny perturbations induced by two-body anisotropy in $S=1,M=0$ channel. In the inset of (d), the red star points show the spectral peak position $\omega_p$ as a function of perturbation strength $\epsilon$; the blue solid line shows the analytical fit $\omega_p\sim\epsilon^{1/3}$.  Here we set $\Omega/\omega_T=0.1$.
%Here the unit of time $t$ (frequency $\omega$) is the inverse of harmonic oscillator frequency (harmonic oscillator frequency) and we set $\Omega=1$.\\
} \label{fig7}
\end{figure}

We have verified above statement in Fig.\ref{fig7}(a), where the Lindblad equation and $H_{eff}$ produce the same dynamics to $\rho_{\uparrow\uparrow}$ in both $\mathcal{PT}$-symmetric phase ($\Gamma<\Omega$) and $\mathcal{PT}$-symmetry broken phase ($\Gamma>\Omega$). In the $\mathcal{PT}$-symmetric phase, $\rho_{\uparrow\uparrow}(t)$ oscillate periodically with time and the corresponding oscillation period and amplitude both increase with $\Gamma$. On contrary, in the $\mathcal{PT}$-symmetry broken phase, $\rho_{\uparrow\uparrow}(t)$ exponentially increases with $t$ and the exponential growth rate
increases monotonously with $\Gamma$ (see Fig.\ref{fig7}(a)).

The next question is how to extract the sensitive spectral response at EP from the dynamics of density matrix (or probability). To do this, one needs to first relate the time-dependent density matrix to the energy splitting at EP. Using Eq.\ref{rho}, we have
\begin{eqnarray}
\rho_{\uparrow\uparrow}(t)
&=&\sum_{j=1}^{3}\sum_{k=1}^{3}\frac{1}{\langle n_j^L|n_j^R\rangle}\langle\uparrow\uparrow|e^{-iH_{eff}(\epsilon)t}|n_j^R\rangle \langle n_j^L|\uparrow\uparrow\rangle \nonumber \\
&\times&\frac{1}{\langle m_k^R|m_k^L\rangle}\langle\uparrow\uparrow|m_k^L\rangle\langle m_k^R|e^{iH^\dag_{eff}(\epsilon)t}|\uparrow\uparrow\rangle \nonumber \\
&=&\Big|\sum_{j=1}^{3}c_j\exp(-i\Delta E_jt)\Big|^2
\label{rho_Heff}
\end{eqnarray}
where $H_{eff}(\epsilon)$ is presented by the $3\times3$ matrix as shown in (\ref{H_3}) and $c_j=\frac{\langle\uparrow\uparrow|n_j^R\rangle \langle n_j^L|\uparrow\uparrow\rangle}{\langle n_j^L|n_j^R\rangle}$, $\Delta E_j=\Delta E\exp(i\frac{2(j-1)\pi}{3})$.
The state vectors $|n_j^R\rangle$, $|m_j^R\rangle$\big($|n_j^L\rangle$, $|m_j^L\rangle$\big) are the eigenstates of $H_{eff}(\epsilon)$\big($H^\dag_{eff}(\epsilon)$\big) and $\langle n_j^{L,R}|=\big(|n_j^{L,R}\rangle\big)^\dagger$, $\langle m_j^{L,R}|=\big(|m_j^{L,R}\rangle\big)^\dagger$.

Note that $\rho_{\uparrow\uparrow}(t)$ is not square integrable function since there exists a positive imaginary part in $\Delta E_2$, i.e., ${\rm Im}(\Delta E_2)=\sqrt{3}\Delta E/2>0$, and therefore it will lead to an exponential growth of $\rho_{\uparrow\uparrow}$ as time $t$ according to Eq.\ref{rho_Heff}.
In order to eliminate the exponential divergence in $\rho_{\uparrow\uparrow}(t)$, we define $\rho'_{\uparrow\uparrow}(t)$ as
\begin{eqnarray}
\rho'_{\uparrow\uparrow}(t)
=\exp\big[-2{\rm Im}(\Delta E_2)t\big]\rho_{\uparrow\uparrow}(t),
  \label{rescale1}
\end{eqnarray}
which is no longer divergent as shown in Fig.\ref{fig7}(b). Since a non-zero constant in time-domian will lead to a delta-function type peak near zero-frequency, in order to obtain clean and crisp signals(frequency distributions) in frequency domain, we should eliminate the non-zero constant in the long time limit and define the rescaled density matrix $\widetilde{\rho}$ as
\begin{eqnarray}
\widetilde{\rho}_{\uparrow\uparrow}(t)&\equiv&\rho'_{\uparrow\uparrow}(t)-\rho'_{\uparrow\uparrow}(t\rightarrow\infty ) \nonumber \\
&=&\exp\big[-2{\rm Im}(\Delta E_2)t\big]\rho_{\uparrow\uparrow}(t) \nonumber \\
&-&\lim_{t\rightarrow\infty}\exp\big[-2{\rm Im}(\Delta E_2)t\big]\rho_{\uparrow\uparrow}(t),  \label{rescale}
\end{eqnarray}
which is square integrable and the corresponding frequency distribution (Fourier transform) $\widetilde{\rho}_{\uparrow\uparrow}(\omega)$ is well-defined.

After the rescaling in Eq.\ref{rescale}, $\widetilde{\rho}_{\uparrow\uparrow}(t)$ becomes an decaying function with oscillation:
\begin{eqnarray}
\widetilde{\rho}_{\uparrow\uparrow}(t)&=&|c_1|^2e^{-\sqrt{3}\Delta Et}+|c_3|^2e^{-2\sqrt{3}\Delta Et}+2Re(c_2^*c_3)e^{-\sqrt{3}\Delta Et}\nonumber \\
&+&2Re\big[c_1c_2^*\exp{(i\frac{3}{2}\Delta Et)}\big]e^{-\frac{\sqrt{3}}{2}\Delta Et} \nonumber \\ &+&2Re\big[c_1c_3^*\exp{(i\frac{3}{2}\Delta Et)}\big]e^{-\frac{3\sqrt{3}}{2}\Delta Et}
\end{eqnarray}
According to the property of Fourier transformation, the type of function $F(t)=\exp(i\omega_0t)\exp(-\omega_1t)$, where $\omega_0$ and $\omega_1$ are all real, will give rise to a peak in frequency domain at location $\omega_0$ and with width $\omega_1$. Hence, after the Fourier transformation, $\widetilde{\rho}_{\uparrow\uparrow}(\omega)$ gets a spectral peak in $\omega=\frac{3}{2}\Delta E$. Given these relations and the fact that $\Delta E\sim\epsilon^{1/3}$ at a third-order EP, one is able to read out such cube-root dependence from the location of the spectral peak of $\widetilde{\rho}_{\uparrow\uparrow}(\omega)$.

In Fig.\ref{fig7}(d), we show the numerical result for $\widetilde{\rho}_{\uparrow\uparrow}(\omega)$ as a function of frequency $\omega$ for different perturbation strength $\epsilon$. Indeed, we can see it has a spectral peak at finite $\omega=\omega_p$ with certain width. In the inset of Fig.\ref{fig7}(d), we further show the extracted $\omega_p$ as a function of $\epsilon$, which indeed show a cube-root dependence $\omega_p\sim\epsilon^{1/3}$.

We propose that the experiment initially prepare two spin-up state $|\uparrow\uparrow\rangle$ at the lowest harmonic level which means $\varrho_{\uparrow\uparrow}(0)\equiv\langle\uparrow\uparrow|\varrho(t=0)|\uparrow\uparrow\rangle=1$ and all other matrix elements equal to zero and then measure the atom numbers of two hyperfine states at the time $t$ as done previously in Ref.\cite{Luo}. The probability of finding two spin-up bosons at the time $t$ in two atoms subspace can be expressed by $\varrho_{\uparrow\uparrow}(t)=N_{\uparrow\uparrow}(t)/N(t)$ where $N(t)$ denotes
%the total times that two atoms is measured experimentally
the total times of measurements and $N_{\uparrow\uparrow}$(t) is the times of finding two spin-up atoms.
After measuring the number of atoms, the rescaled probability $\widetilde{\rho}_{\uparrow\uparrow}(t)$ is acquired by applying the transformations in (\ref{density_matrix}), (\ref{rescale1}) and (\ref{rescale}). Finally, we can obtain the frequency distribution $\widetilde{\rho}_{\uparrow\uparrow}(\omega)$ through simple Fourier transform algorithm which can reflect the features of energy splitting at EP.

In above we have proposed to detect the ultra-sensitive energy splitting by measuring the spectral distribution of the rescaled probability. It is worth pointing out that the above proposal does not rely on the choice of initial density matrix or the choice of dynamics in particular spin space, as long as they are in a given particle number sector. This is because the energy splitting of eigenstates will manifest themselves in the dynamics of any physical quantities.

In the last part of this section, we discuss the experimental relevance to the two-body and three-body interaction anisotropies. We consider the spin-1 $^{87}$Rb atoms
whose effective interaction in the mean-field limit can be written as $U(\mathbf{r})=(c_0+c_2\mathbf{F_1\cdot F_2})\delta(\mathbf{r})$, where $c_0=4\pi\hbar^2(a_0+2a_2)/3m$, $c_2=4\pi\hbar^2(a_0-a_2)/3m$. Since the scattering lengths $a_0$ and $a_2$ in $^{87}$Rb atoms are very close ($a_0=101.8r_{{\rm Bohr}}$, $a_2=100.4r_{{\rm Bohr}}$)\cite{Rb87} and can be finely tuned by changing the magnetic field\cite{Greene1}, the spin-dependent interaction $c_2\mathbf{F_1\cdot F_2}$ can serve as perturbation.
As discussed in Ref.\cite{Pan}, a two-species Bose gas with nearly spin-independent interaction can be achieved by using the lowest two hyperfine states, i.e., $|\hspace{-0.11cm}\uparrow\rangle=|F=1, m_F=0\rangle$ and $|\hspace{-0.11cm}\downarrow\rangle=|F=1, m_F=-1\rangle$ and meanwhile, the third hyperfine state $|F=1, m_F=1\rangle$ is eliminated adiabatically by the finite quadratic Zeeman energy. In this case, a two-body interaction anisotropy $\epsilon=(a_0-a_2)/3$ is generated in $|\hspace{-0.11cm}\uparrow\uparrow\rangle$ scattering channel.

To realize the three-body interaction, one can tune the magnetic field around an Efimov resonance in particular collision channel, e.g., $|F=1, m_F=-1\rangle$ channel\cite{Cornell} or $|F=1, m_F=+1\rangle$ channel\cite{Marte,Smirne} in $^{87}$Rb
atoms. One can also utilize transverse confinement to generate non-negligible three-body strengths\cite{Mazets, Pricoupenko, Nishida, Guijarro}.
In the similar scheme, a three or many more species Bosonic system and the corresponding interaction anisotropies can be realized in experiment.

\section{Summary}\label{sec7}
In summary, we have demonstrated the properties of higher-order EPs and their associated spectral sensitivities due to the interplay between $\mathcal{PT}$-symmetric potential and particle interactions in a 3D trapped boson system with arbitrary spin.
%In summary, we have studied the interplay of interaction and $\mathcal{PT}$-symmetric potential to the property of high-order EPs and their associated spectral sensitivities in a 3D trapped boson system with an arbitrary spin.
Consistent with our previous work on two-species bosons in 1D\cite{Pan}, here we show that a 3D Bose gas in the repulsive scattering branch can also be used to create high-order EPs. We have exactly solved the non-Hermitian two-body problem for spin-1/2 and spin-1 bosons in a 3D harmonic trap, and verified the mean-field treatment  in predicting the eigen-spectrum of repulsive branch in weak coupling regime. We further utilize the mean-field treatment to study the properties of EPs for a small cluster of spin-1/2 and spin-1 bosonic systems (see Table \ref{table1}), and then generalize to many-body systems with arbitrary spin. Finally, we put forward an experimental proposal to detect the ultra-sensitive spectral response at high-order EPs. Our main conclusions are summarized as follows:

(I) For $N$ spin-$s$ bosons under $\mathcal{PT}$ potential (\ref{H_PT}), a $(2Ns+1)$-th order EP will occur at $\Omega=\Gamma$ in the presence of a spin-independent interaction where all the eigen-energies and all eigen-vectors coalesce into a single energy and a single vector.

(II) Based on the high-order EP created above, when a tiny interaction anisotropy in the $k$-body collision sector is turned on, the original EP will split into $[\frac{2Ns+1}{2ks+1}]$ groups of sub-EPs, and the most sensitive spectral splitting scales as $\epsilon^{\frac{1}{2ks+1}}$.  We have numerically verified this conclusion for a small cluster of spin-1/2 and spin-1 boson systems.

(III) The spectral response at the high-order EP can be detected in cold atoms experiment by measuring the dynamics of density matrix (probability) in given particle number sector. After the Fourier transformation of rescaled dynamics, one can extract the ultra-sensitive spectral response from the probability peak in frequency domain.

These results reveal the intriguing interplay effect between interaction, non-Hermitian potential and the bosonic statistics. The phenomenon of sensitive spectral response may be detected through the spectroscopy measurement in the s-wave scattering Bose gas. Based on these results, a powerful atomic sensors may be designed in spinor Bose gases with tunable few-body forces.\\

\begin{acknowledgments}
The work is supported by the National Key Research and Development Program of China (2018YFA0307600, 2016YFA0300603), and the National Natural Science Foundation of China (No.11622436, No.11425419, No.11421092, No.11534014).
\end{acknowledgments}

%\newpage


\begin{thebibliography}{99}
\bibitem{footnote} By this definition, the relation $\sum_j |\mu_j^R\rangle \langle \mu_j^L|/\langle \mu_j^L|\mu_j^R\rangle=\Bbb1$ is always true except at EPs($\Gamma=1$).
\end{thebibliography}
\end{document}